\documentclass[10pt]{iopart}
\usepackage{amssymb,graphicx}
\expandafter\let\csname equation*\endcsname\relax
\expandafter\let\csname endequation*\endcsname\relax
\usepackage{amsmath}

\usepackage{url}
\usepackage{color}
\begin{document}

\title[Symmetry analysis of trimer-based all-dielectric metasurfaces] {Symmetry analysis of trimer-based all-dielectric metasurfaces with toroidal dipole modes}

\author{Victor Dmitriev$^1$, Anton S Kupriianov$^{2}$\footnote{Present address: College of Physics, Jilin University, 2699 Qianjin Street, Changchun 130012, China.}, Silvio Domingos Silva Santos$^{1}$, and Vladimir R Tuz$^{3}$}
\address{$^1$Electrical Engineering Department, Federal University of Para, PO Box 8619, Agencia UFPA, CEP 66075-900 Belem, Para, Brazil}
\address{$^2$ College of Physics, Jilin University, 2699 Qianjin Street, Changchun 130012, China}
\address{$^3$ State Key Laboratory of Integrated Optoelectronics, College of Electronic Science and Engineering, International Center of Future Science, Jilin University, 2699 Qianjin Street, Changchun 130012, China}
\ead{kupriianov1116@mails.jlu.edu.cn}


\begin{abstract}
Herein, we discuss the conditions for excitation of symmetry-protected toroidal dipole modes in all-dielectric metasurfaces composed of trimer or twin-trimer clusters of dielectric disks. Such metasurfaces permit enhanced light-matter interaction due to spatially confined light in resonant systems with a high-quality factor. To describe characteristics of toroidal modes existing in the clusters, we use the magnetic dipole moments approximation, group-theoretical methods, group representation theory, symmetry-adapted linear combination method, and circuit theory. To validate the obtained theoretical results, we fulfill both full-wave numerical simulations and microwave experiments. In particular, we have shown that the toroidal dipole mode appears as a quasi-dark state of the trimer. It can be excited in the metasurface by the field of a linearly polarized wave, providing the symmetry of the trimer is properly reduced. In the metasurface, the properties of the toroidal dipole mode are determined primarily by the parameters of a single trimer and are not a consequence of the periodicity of the array. The coupling of the toroidal dipole modes in the twin-trimers can appear in both bonding and anti-bonding fashion resulting in different orders of the net toroidal dipole moment. Due to the unique field configuration of these modes, the proposed metasurfaces can be considered as a platform for efficient light-matter interaction for enhanced absorption, nonlinear switching, and sensing.

\end{abstract}
%
\vspace{2pc}

%
\submitto{\JPD}
%
%
\ioptwocol
%

\section{Introduction}
\label{sec:Intro}
Macroscopically, an electric dipole originates from the separation of positive and negative charges, while a magnetic dipole is generated by a closed loop of electric currents. If several magnetic dipoles are arranged into a closed loop in a head-to-tail fashion, the polar toroidal moment arises \cite{zeldovich_JETP_1957, Artamonov_JETP_1985, Dubovik_JETP_1986, Dubovik_PhysRep_1990, Afanasiev_JPhysA_1995}. This toroidal moment is characterized by a vortex distribution of magnetic dipoles and this state is asymmetric under both the reversal of time and the reversal of space \cite{Papasimakis_NatMater_2016}. The induction of the \textit{static} toroidal moments in the matter is referred to as the toroidicity. In multiferroic materials the ferrotoroidicity is the fourth kind of ferroic order after ferromagnetism, ferroelectricity, and ferroelasticity \cite{Loidl_JPhysCondensMatter_2008} (in what follows, the prefix `ferro' for the ferrotoroidicity is be omitted). Recently, multiferroic materials with the toroidicity have been widely studied in the context of molecular, condensed matter, and applied physics. It is expected that they may find application in devices for advanced magnetic data storage and quantum computations (for review on the appearance of the toroidicity in natural materials, see references \cite{Gnewuch_JSolidStateChem_2019, Li_DaltonTrans_2019}).

Natural multiferroic materials bearing toroidal states are scarce, and the manifestation of the toroidicity in such materials is usually weak. Therefore, special attention is paid to the implementation of the toroidicity in artificial media such as metamaterials \cite{Zheludev_Science_2010}. The \textit{dynamic} toroidal moments can be induced by oscillating electromagnetic fields in metamaterials when their constituent `meta-atoms' are properly designed. In this case, the toroidal response can be engineered to be sufficiently strong. Traditionally, the toroidicity is sought in metamaterials composed of clusters (`meta-molecules') consisting of four or more subwavelength particles, which can be made of either metallic \cite{Guo_2014, Zheludev_PhysRevB_2016} or dielectric materials \cite{Zheludev_PhysRevX_2015, Tasolamprou_PhysRevB_2016}, or their combination \cite{Tang_IEEEPhotonJ_2016}, depending on the desired operating frequency range. Such systems permit an enhanced light-matter interaction due to spatially confined light in resonant elements with high-quality (high-$Q$) factor and can be used as sensors \cite{Gupta_ApplPhysLett_2017}, narrowband filters, switches \cite{Gupta_AdvMater_2017}, modulators, multifunctional on-chip lasers \cite{Zheludev_SciRep_2013}, and nonlinear components (for review on application of metamaterials bearing toroidal moments see reference \cite{Talebi_Nanophoton_2018}).

The spatial feature of the toroidal states is straightforwardly related to three-dimensional (3D) structures (metamaterials) composed of clusters with circular arrangement of subwavelength particles. When the operating frequency of a metamaterial increases, the size of the corresponding constituent particles decreases, and the implementation of such a metamaterial becomes cost-inefficient. Therefore, there is a need to develop two-dimensional (2D) platforms (metasurfaces) with highly optimized meta-molecule designs supporting toroidal moments \cite{Gupta_RevPhys_2020}. Apart from a variety of known metasurface designs supporting toroidal modes, we further distinguish a particular class of resonant metasurfaces which allow to obtain the strongest resonant response due to excitation of the symmetry-protected toroidal states attributed to the so-called trapped (quasi-dark) modes \cite{Fedotov_PhysRevLett_2007, khardikov_JOpt_2010, Tuz_OptExpress_2018}.

In particular, the toroidal mode existing in metasurfaces is considered as symmetry protected state, when it cannot be excited by normally incident waves without the symmetry breaking of the metasurface unit cell \cite{Khardikov_JOpt_2012, He_PhysRevB_2018, Kupriianov_PhysRevApplied_2019}. The symmetry breaking is realized by introducing a conscious perturbation into the meta-molecule design. This perturbation transforms an inherently non-radiative state into a weakly radiative one producing the mode coupling with incoming radiation. This coupling results in the arising narrow resonance in the metasurface spectra. 

Moreover, due to low inherent losses in constituent materials, the symmetry-protected toroidal modes in all-dielectric metasurfaces benefit from very high-$Q$ factor of corresponding resonances and strong near-field confinement. To realize the excitation of toroidal modes, several metasurface cluster designs have been recently proposed. The manifestation of toroidal modes is revealed to be possible in meta-molecules composed of a single particle \cite{Zografopoulos_Nanomater_2018, Zografopoulos_AdvOptMater_2019}, dimer \cite{Zhang_OptLett_2018}, trimer \cite{Xu_AdvOptMater_2019}, quadrumer \cite{Tuz_ACSPhotonics_2018, Sayanskiy_AdvOptMater_2018, Zhang_2019}, pentamer and hexamer \cite{Zhang_AdvTheorySimul_2019}, as well as rhomboid clusters \cite{Kupriianov_PhysRevApplied_2019}. Note, that the multi-element cluster designs benefit from additional flexibility in control of the toroidal mode excitation proceeding to multi-dimensional parameter space. 

When analyzing electromagnetic characteristics of isolated dielectric particles with high structural symmetry, the analytical approach of multipole expansion is widely used to derive the fields induced by incoming radiation inside and outside of the particles \cite{Afanasiev_Nuclei_1998, Gurvitz_LPOR_2019}. For the study of mechanisms of the toroidal mode excitation in meta-molecules composed of several particles with broken symmetry, similar rigorous analytical approaches are hardly possible because of the geometric complexity of the clusters. 

Different numerical methods can be useful in this case. For analysis of the structures with quasi-dark modes, along with numerical methods, some valuable information about the characteristics of electromagnetic fields can be obtained by using semi-analytical techniques, such as the method of complete multipole expansion accounting for the toroidal modes \cite{Dubovik_PhysRep_1990} and also approximate models, where the processes are described by simple differential or algebraic equations. Among the approximate models, one can mention the classical model of coupled harmonic oscillators \cite{Joe_PhysScr_2006, Satpathy_EurJPhys_2012, Tuz_PhysScr_2015, Gallinet_Chapter_2018} and the electrical ${\cal RLC}$ circuits method \cite{Lv_SciRep_2016, Harden_EurJPhys_2011}. Besides, the characteristics of fields in the particle clusters can be modeled in the discrete dipole approximation \cite{Evlyukhin_PhysRevB_2011}. Still another appropriate approach is the temporal coupled-mode theory (TCMT) \cite{Fang_Chapter_2018, Fan_JOptSocAmA_2003, Souza_OptExpress_2016}.

Group-theoretical methods based solely upon the symmetries of their underlying constituent particles are very effective for revealing general electromagnetic properties of metasurfaces \cite{Padilla_OptExpress_2007, Dmitriev_Metamat_2011, Dmitriev_IEEEAntennas_2013, Hopkins_PhysRevA_2013, Dmitriev_JApplPhys_2019, Sadrieva_PhysRevB_2019}. From the transformation properties of an electromagnetic basis under symmetries of the particles, one can determine the natural modes (eigenwaves) of the particles and their collective behaviors responsible for the resonant characteristics of the entire metasurface. These methods include the point group and representation theories \cite{hamermesh_book_1962, Bradley_book_2009}. When the metasurfaces are modeled in the discrete dipole approximation, the symmetry-adapted linear combinations (SALC) method can be involved. This method combines the individual modes of particles forming the cluster by using irreducible representations (IRREPs) of the corresponding symmetry group. The group-theoretical methods permit defining the necessary conditions for the existence of symmetry-protected modes in the metasurfaces \cite{Overvig_PhysRevB_2020}. These modes are recently referred to as the symmetry protected bound states in the continuum (BICs) \cite{Shabanov_PhysRevLett_2008, Bulgakov_PhysRevB_2008, Plotnik_PhysRevLett_2011, Koshelev_PhysRevLett_2018}.

In our study, we propose a generalized theory describing characteristics of a toroidal dipole mode, which is implemented in cluster-based all-dielectric metasurfaces. In particular, we discuss general conditions for the toroidal mode existence in trimer and twin-trimer clusters forming the metasurface, and propose a way to access this mode with the field of a normally incident linearly polarized wave. Our theoretical description involves the magnetic dipole moments approximation, group-theoretical methods, group representation theory, symmetry-adapted linear combination method, and circuit theory. Obtained theoretical findings are checked against numerical simulations performed with the commercial COMSOL Multiphysics solver and experimental measurements for the metasurface prototypes designed to operate in the microwave range.
\section{\label{sec:statement}Outline of the problem} 
In what follows, we consider a metasurface composed of dielectric disk-shaped particles. The disks are distributed in the $x$-$y$ plane, forming an infinitely large two-dimensional (2D) array. We assume that the main building block of the array consists of three disks (i.e., it is a trimer). The disks are located in the trimer at the vertices of an equilateral triangle. We perform a comparative study of characteristics of two particular designs of the metasurface. For the first design, trimers are disposed into a square unit cells, while for the second design, two trimers are ordered into a twin-trimer cluster to construct either rectangular or rhomboid unit cells of the metasurface. Drawing analogy with molecular physics, the hierarchy of the constituent elements of the metasurface in our study can be established as follows: the dielectric disk can be considered as a meta-atom, while the trimer and twin-trimer can be associated with a meta-molecule and meta-macromolecule, respectively. 

\begin{figure}[t!]
\centering
\includegraphics[width=1.0\linewidth]{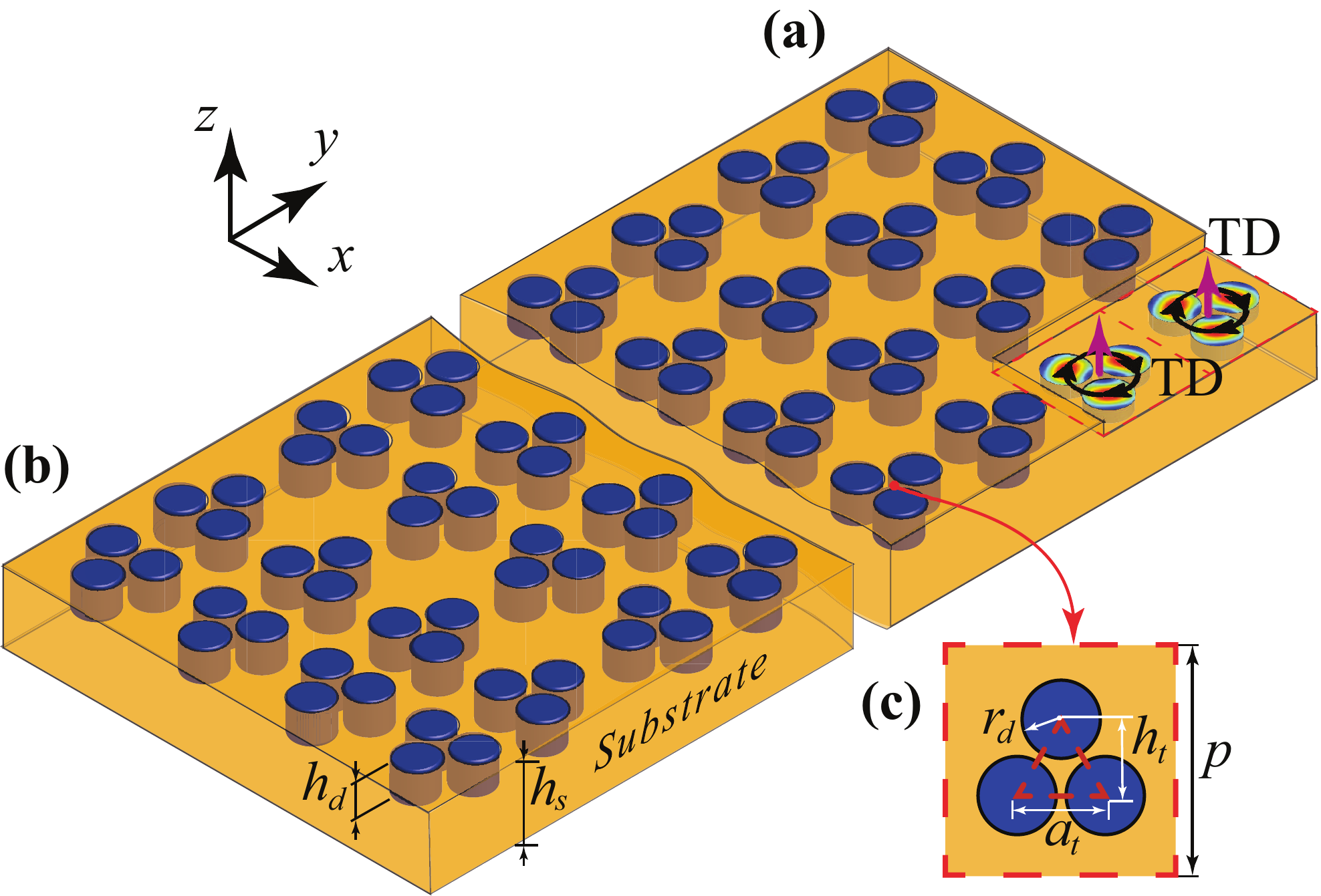}
\caption{A schematic view of two designs of an all-dielectric metasurface composed of (a) trimers and (b) twin-trimers, and (c) a sketch of the unit cell with a trimer.} \label{fig:fig1}
\end{figure}

This hierarchy in the metasurface composition can be understood from the schematic representation shown in figure~\ref{fig:fig1}. Here we introduce the notations that $r_d$ and $h_d$ are the radius and thickness of the disks, and $a_t$ and $h_t=\sqrt{3}a_t/2$ are the base and height of the equilateral triangle forming the trimer, respectively. The cluster size is $p$. The disks are made of a nonmagnetic dielectric material with relative permittivity $\varepsilon_d$. They are immersed symmetrically (preserving the plane of symmetry $z=0$) into a dielectric host with relative permittivity $\varepsilon_s$ and thickness $h_s$.

We suppose that the metasurface under consideration is illuminated by a normally incident ${\bf k} = \{0,0,k_z\}$ linearly polarized wave. The wavelength of the incident wave is $\lambda=2\pi c/\omega$, where $c$ is the light velocity in free space and $\omega$ is the angular frequency. In general, the excitation of a resonant system can be provided by different components of the incident electromagnetic field. In our case, it is convenient to discuss the problem using the magnetic field of the incident wave as an excitation source. By definition, the in-plane symmetry of the disk is $C_{\infty v}$ (in the Sch\"oenflies notation \cite{Barybin_book_2002}) and the in-plane symmetry of the magnetic field $\bf H$ of the incident wave is $C_{s}$ with the vector $\bf H$ perpendicular to the plane of symmetry.

For an informative description of the symmetry of a periodic array forming the metasurface, one should consider the scheme of symmetry points of an extended Brillouin zone in the reciprocal space \cite{Overvig_PhysRevB_2020}. This can give a complete description of the symmetry properties of the metasurface. However, for simplicity, in our subsequent analysis, we restrict ourselves to discussing metasurfaces only by point symmetries of their unit cells, excluding from consideration the periodicity of the structure. Thus, we consider the $\Gamma$-points in the first Brillouin zone, which corresponds to the lowest resonant frequencies.
\section{\label{sec:isodisk}Magnetic dipole resonance in isolated dielectric disk}
A dynamic toroidal magnetic dipole can be approximately interpreted as a state produced by a set of magnetic dipoles, which are azimuthally distributed in space \cite{Dubovik_JETP_1986}. Each magnetic dipole originates from a closed loop of displacement currents induced in the individual (isolated) particles by the fields of incident electromagnetic wave. Here our goal is to distinguish a corresponding mode of the isolated cylindrical dielectric resonator (disk) supporting necessary resonant conditions for the existence of the magnetic dipole moment oriented parallel to the metasurface plane (the $x$-$y$ plane in the chosen coordinate frame; see figure \ref{fig:fig1}). 

\begin{figure}[t!]
\centering
\includegraphics[width=0.8\linewidth]{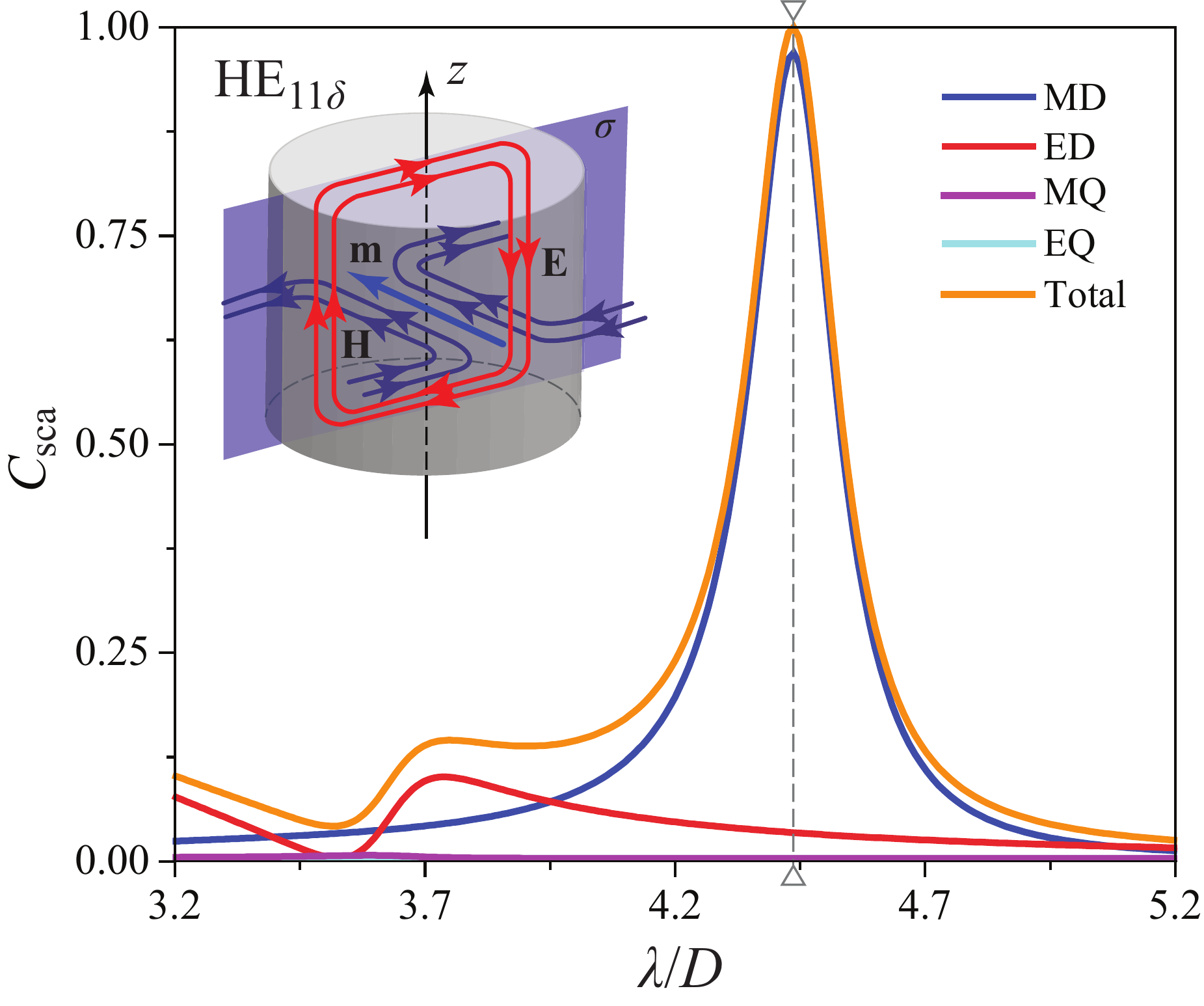}
\caption{Normalized contributions of the four lowest order multipole moments [magnetic dipole (MD), electric dipole (ED), magnetic quadrupole (MQ), and electric quadrupole (EQ)] to the scattering cross-section of a single dielectric disk frontally irradiated by a linearly polarized wave. Schematic view of the electric $\bf{E}$ and magnetic $\bf{H}$ fields distribution of the HE$_{11\delta}$ mode and orientation of the magnetic dipole moment $\bf{m}$ are given in the inset. The thickness-to-diameter ratio and permittivity of the disk are $h_d/D = 5/8$ and $\varepsilon_d=23$, respectively. The disk is disposed in free space ($\varepsilon_s=1$).} \label{fig:fig2}
\end{figure}

The contribution of the lowest-order multipoles to the scattering cross section is shown in figure~\ref{fig:fig2} for the disk frontally irradiated (i.e., along the $z$ axis) by a linearly polarized wave. These multipoles are calculated in a standard way \cite{Jackson_book_1998}. For our calculations, we consider that the disk is made of a lossless high-$\varepsilon$ ceramic material. We use this material in view of our experimental capabilities and techniques \cite{Sayanskiy_PhysRevB_2019, Kupriianov_PIERS_2019}. Nevertheless, our results can be easily scaled for the disks with others geometrical and material parameters operating in the desired spectral range. Therefore, in what follows we normalize all geometrical parameters of the problem on the disk diameter $D = 2 r_d$. Unless otherwise stated, in our numerical study we suppose that the disk is disposed in free space.

In the chosen spectral range, the strongest resonance in the scattering cross-section arises from the magnetic dipole attributed to the manifestation of the lowest hybrid HE$_{11\delta}$ mode \cite{Mongia_1994} of the cylindrical resonator (HEM$_{11}$ mode in the nomenclature of reference \cite{Kishk_book_2003}). This resonance is marked out on the wavelength scale by the grey arrows. For the HE$_{11\delta}$ mode, the fields are azimuthally dependent and the $H_z$ component is sufficiently smaller than the $E_z$ component. For this mode, the main contribution to the scattering cross-section is from the transversely oriented magnetic dipole (see figure \ref{fig:fig2}). 

The resonant frequency $\omega$ of the magnetic dipole resonance can be approximated by
\begin{equation}
\omega \approx 2\cdot1.841 c(Dn)^{-1}, 
\label{eq:2}
\end{equation}
where $c$ is the light velocity and $n=\sqrt\varepsilon_d$ is the refractive index of the material from which the disk is made. Equation (\ref{eq:2}) corresponds to the first resonance of an infinitely extended dielectric cylinder, whose lateral surface is substituted by a perfectly magnetic conductor (TM$_{11}$ mode, $J_1^{\prime}(x)=0$, $x=1.841$ \cite{Watkins}). This equation is only a very rough approximation which does not take into account for the thickness of the disk and the fields around it. Nevertheless, this simple analytical expression is useful for our subsequent estimations.

\begin{figure*}[ht!]
\centering
\includegraphics[width=1.0\linewidth]{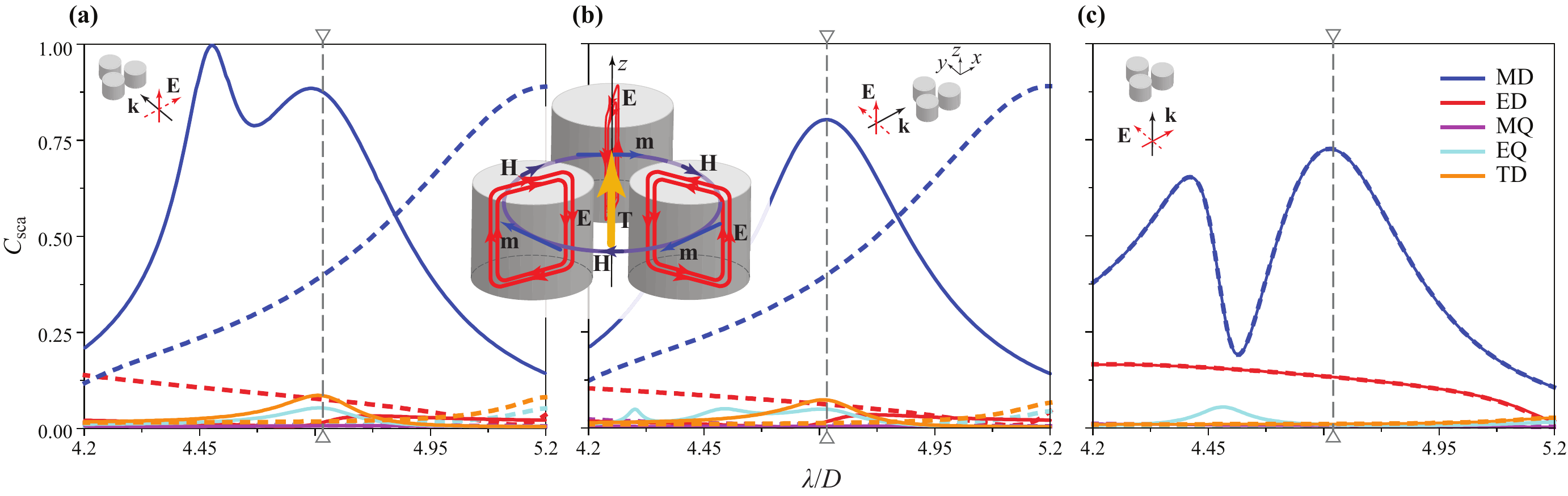}
\caption{Same as in figure \ref{fig:fig2} but for a trimer which is (a), (b) laterally and (c) frontally irradiated by a linearly polarized wave, with accounting for the toroidal dipole (TD) term in the multipole decomposition. Solid and dashed lines represent scattering for two different orthogonal polarizations. Schematic view of the electromagnetic field distribution of the collective mode of the trimer and orientation of the polar toroidal dipole moment $\bf T$ are given in the inset. All parameters of the disks are the same as in figure \ref{fig:fig2} and $a_t/D=9/8$.}
\label{fig:fig3}
\end{figure*}

Considering small dimensions of the disk with respect to the wavelength in free space, the fields related to this mode can be presented in quasi-static approximation. In linear regime, the moment $\bf m$ of the magnetic dipole is proportional to the incident magnetic field $\bf H$ as follows: 
\begin{equation}
{\bf m} =\alpha \bf H,
\label{eq:magnetic}
\end{equation}
where $\alpha$ is the magnetic polarizability. This polarizability can be obtained by the methods described elsewhere \cite{Bohren_book_1998, Merchiers_PhysRevA_2007}. 

The 2D symmetry of the magnetic dipole is $C_{s}$ with the dipole axis oriented perpendicularly to the plane of symmetry $\sigma$ (see the inset in figure~\ref{fig:fig2}), which behaves as a magnetic wall, whereas the plane of symmetry $z = 0$ behaves as an electric wall (the transformation properties of both the magnetic field $\bf H$ and magnetic dipole moment $\bf m$ are presented in equation (\ref{eq:A1}) of \ref{sec:appendixA}).
\section{\label{sec:isotrimer}Toroidal dipole mode in trimers} 
\subsection{General description of trimer} 

A polar toroidal dipole may arise in different types of meta-molecules having a rotational arrangement of their meta-atoms \cite{Zhang_AdvTheorySimul_2019}. This toroidal state appears from a circular distribution of the corresponding magnetic dipoles induced in particles forming the cluster. It is associated with the collective hybrid mode. 

It is revealed \cite{Tasolamprou_PhysRevB_2016} that the spectral separation of the toroidal dipole against the other higher-order electric and magnetic multipoles existed in the meta-molecule is more efficient in the clusters composed of the odd number of the constituent particles. This conclusion is obtained by considering the 2D problem on the existence of toroidal states in an isolated trimer consisting of infinitely extended cylinders. With this information, here our goal is to identify the conditions for the appearance of a polar toroidal dipole in a trimer-based cluster consisting of disks. Such a cluster provides the simplest rotational arrangement of the particles able to support toroidal states \cite{Xu_AdvOptMater_2019}.

When dielectric disks are combined into a cluster, the overall meta-molecule response is determined mainly by the near-field coupling between the particles inside the clusters rather than by the electric and magnetic dipole modes of individual disks. The out-of-plane toroidal dipole moment arises from a collective mode originating from the excitation of three magnetic dipole moments, where each magnetic moment belongs to a particular resonator forming the cluster. The electromagnetic field distribution in each resonator corresponds to the characteristics of the HE$_{11\delta}$ mode. Under such conditions, each resonator supports the dipole moment $\bf m$ located in-plane (i.e., in the $x$-$y$ plane). In the collective mode, the magnetic moments of the three resonators are oriented along the circumference passing through the centers of the resonators (see the inset in figure \ref{fig:fig3}). Their combined effect leads to the appearance of a closed loop of the magnetic field within the trimer, which potentially may generate a toroidal dipole moment oriented out-of-plane (i.e., along the $z$ axis).

To corroborate the existence of a toroidal dipole in the trimer, we employ multipole decomposition accounting for the toroidal dipole term to quantify the contribution of each multipole to the scattering cross-section. In the multipole decomposition \cite{Savinov_PhysRevB_2014}, the toroidal dipole moment is defined as
\begin{equation}
{\bf T} = \frac {1}{10c} \int \left[({\bf r}\cdot {\bf J})\cdot{\bf r}-2r^2 {\bf J}\right] d^3r,
\label{eq:toroid}
\end{equation}
where $J_z=\rmi\omega\varepsilon_0(\varepsilon_d-1)E_z$ is the dominant component in the polarization current $\bf J$ and $\bf r$ is the position vector.  In our simulation, both the lateral and the frontal irradiation of the trimer by a linearly polarized wave are investigated. For each geometry, two possible orthogonal polarizations of the incident wave are considered. The results of the multipole decomposition are summarized in figure \ref{fig:fig3}. The solid and dashed lines refer to the scattering cross-section obtained for the trimer irradiation by waves with different polarization (see the designation of the vector $\bf E$ in the figure insets).

Under the lateral irradiation of the trimer, when the vector $\bf E$ is oriented along the $z$ axis, a maximum appears in the curves related to the magnetic dipole, electric quadrupole, and toroidal dipole at the wavelength which is close to the resonant wavelength fixed above in figure \ref{fig:fig2} for the HE$_{11\delta}$ mode. As before, this resonant frequency is marked out on the wavelength scale by the grey arrows. At this wavelength, the electric dipole has a local minimum, whereas the contribution of the magnetic quadrupole has no resonant feature [see solid lines in figures \ref{fig:fig3}(a) and \ref{fig:fig3}(b)]. Contrariwise, the resonant feature appears only for the magnetic dipole under the frontal irradiation of the trimer by the wave of both polarizations [see figure \ref{fig:fig3}(c)]. 

Therefore, it is revealed that a toroidal moment is observed only in the case when the trimer is irradiated laterally, although it is somewhat masked by the electric quadrupole. Due to symmetry of the trimer, this toroidal moment cannot be excited under the frontal irradiation, since the initial dipole magnetic moments are compensated along the $x$ and $y$ axis in this case. For such a geometry, the polar toroidal dipole turns out to be a dark state for the fields of a linearly polarized wave \cite{Gurvitz_LPOR_2019, Miroshnichenko_NatCommun_2015}. Nevertheless, as we show below, it is possible to access this state in the metasurface under the frontal irradiation conditions by reducing trimer symmetry.
\subsection{Symmetry analysis of trimer eigenmodes} 
In the 2D symmetry (i.e., in the $x$-$y$ plane for the given problem geometry), the highest group of symmetry of the trimer whose identical disks are disposed at the vertices of an equilateral triangle is $C_{3v}$. This group consists of elements $C_{3}$ and $C_{3}^{-1}$ describing the anti-clockwise and clockwise rotations on $2\pi/3$ around the triangle center, respectively, and contains two 1D representations $A$ and $B$, and one 2D representation $E$ (see IRREPs of the $C_{3v}$ group in table B1 of \ref{sec:appendixB}). Three vertical planes of symmetry of the trimer are denoted as $\sigma_1$, $\sigma_2$, and $\sigma_3$. They pass through the center and the vertices of the triangle (see figure \ref{fig:figC1} in \ref{sec:appendixC}). 
 
The structure of the dipole moments for the eigenmodes of the trimer belonging to the $C_{3v}$ symmetry is calculated involving the SALC method. In general, there are six eigenmodes in the cluster. The polar toroidal mode belongs to IRREP $A$, the radial mode belongs to IRREP $B$, and two degenerate dipole modes and two degenerate quadrupole modes belong to IRREP $E$ (see table C1 in \ref{sec:appendixC}). In the problem under consideration, all eigenmodes are separated and mutually orthogonal due to the rigorous group-theoretical approach. Among these eigenmodes, only the dipole modes $D_x$ and $D_y$ can be excited in the trimer by a linearly polarized wave under the frontal irradiation conditions.

All six eigenmodes of the trimer are also identified in reference \cite{Tasolamprou_PhysRevB_2016} from numerical study (see figure 5 in this reference), where some discrepancies with our results are noticed. These discrepancies are explained by a specific excitation  (lateral) of the trimer in reference \cite{Tasolamprou_PhysRevB_2016}, while our consideration is related to eigenstates. 

\subsection{Reducing trimer symmetry} 
In the frontal geometry of the problem, the polar toroidal dipole is a symmetry-protected state (i.e., it behaves like a dark mode). Access to such a state can be realized by reducing trimer symmetry. The $C_{3v}$ group has two subgroups, namely, $C_{3}$ and $C_{s}$ (the correspondence of IRREPs of the $C_{3v}$, $C_{3}$, and $C_{s}$ groups is given in table B2 of \ref{sec:appendixB}). Thus, we study the characteristics of the trimer whose symmetry is reduced to either $C_{3}$ or $C_{s}$ group, revealing the conditions for the polar toroidal dipole excitation. 

\begin{figure}[t!]
\centering
\includegraphics[width=1.0\linewidth]{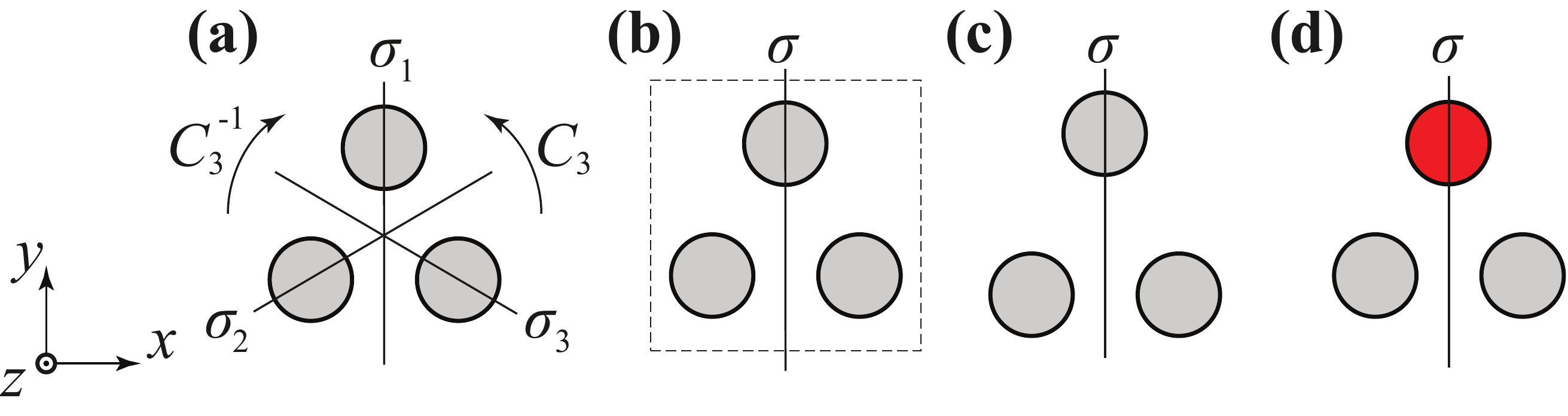}
\caption{Possible ways to reduce the trimer symmetry by transformations related to the $C_{3}$ and $C_{s}$ groups, (a) elements of symmetry of trimer, (b) perturbation by symmetry of the square unit cell, (c) dislocation of one of the disks, and (d) substitution of a disk with other material or geometrical parameters (this perturbed disk is marked by red color).}
\label{fig:fig4}
\end{figure}

Transformations related to the $C_{3}$ group imply the rotation of the trimer around its axis as shown in figure \ref{fig:fig4}(a). It is obvious that such transformations do not resolve the given problem of the toroidal dipole excitation. Thus, only the $C_{3v}$ symmetry reduction to the $C_{s}$ group may provide the desired result. Under such a reduction, the IRREPs $A$ and $B$ of the $C_{3v}$ group degenerate into the IRREPs $A_1$ and $B_1$ of the $C_{s}$ group, respectively (see table B2 of \ref{sec:appendixB}).

A straightforward way to reduce the trimer symmetry is a deposition of a cluster of three identical disks with the $C_{3v}$ symmetry into a square unit cell with the $C_{4v}$ symmetry when constructing metasurface, as shown in figure \ref{fig:fig4}(b). Remarkably, for such a symmetry reduction, it was experimentally confirmed \cite{Xu_AdvOptMater_2019} that a metasurface composed of trimers arranged in square unit cells indeed maintains a toroidal response being excited by a normally incident linearly polarized wave with a proper polarization. 

\begin{figure*}[t!]
\centering
\includegraphics[width=0.7\linewidth]{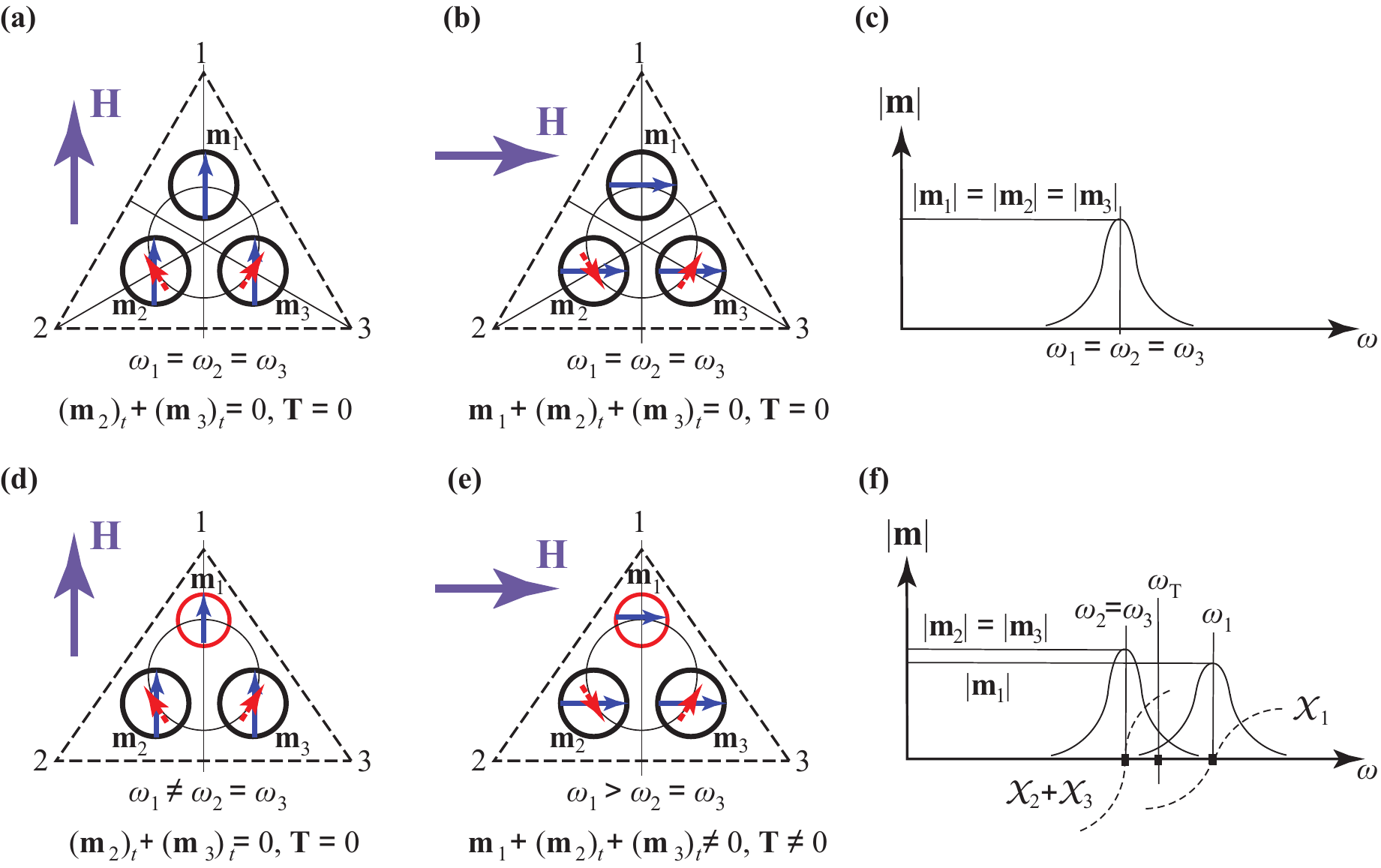}
\caption{Schematic illustration of the toroidal dipole excitation in a trimer belonging to (a)-(c) the $C_{3v}$ group and (d)-(f) the $C_s$ group. Here $\omega_i$ ($i=1,2,3$) is the resonant frequency of the HE$_{11\delta}$ mode in the corresponding resonator, $\omega_\textrm{T}$ is the frequency of the collective (toroidal) mode, and the subscript $t$ means the magnetic dipole moment projection on a circle, i.e., it is related to the tangential components of $\bf m$.}
\label{fig:fig5}
\end{figure*}

Another kind of perturbation can be performed by changing the arrangement of particles in the trimer while fixing all the disks identical and keeping one of the planes of symmetry $\sigma$ unchanged (i.e., this perturbation transforms the equilateral triangle forming the trimer into an isosceles one). The resulting perturbed trimer is presented in figure \ref{fig:fig4}(c). This reduces the resulting symmetry of the meta-molecule to the $C_{s}$ group. 

The $C_{3v}$ group can also be reduced to its $C_{s}$ subgroup by replacing one disk in the trimer by a resonator having a different diameter, shape, or made of a material whose permittivity is different from that of the remaining two disks in the trimer. Such perturbations can be classified as in-plane symmetry breaking. Accordingly, a violation of the thickness of one of the disks in the trimer is related to out-of-plane symmetry breaking \cite{Kupriianov_PhysRevApplied_2019}. The resonator perturbed either in-plane or out-of-plane is marked in figure \ref{fig:fig4}(d) by red color. Both these types of perturbation manifest themselves in a shift of the corresponding resonant frequency. One can estimate this frequency shift by using equation (\ref{eq:2}).

\subsection{Conditions for toroidal mode excitation}

A resonant system coupling with an incident wave depends on several factors, in particular, on polarization of the incident wave and on the angle of its incidence. In accordance with the optical theorem \cite{Overvig_PhysRevB_2020, Jackson_book_1998}, the coupling coefficient $\gamma_m$ of the incident wave with $m$-th mode of the resonant system is proportional to the integral
\begin{equation}
\gamma_m \propto \int \!\!\int[{\bf E}_{i}^{\ast}\times{\bf H}_{m}+{\bf E}_{m}\times {\bf H}^{\ast}_{i}]dxdy,
\label{eq:3}
\end{equation}
where ${\bf E}_m$ and ${\bf H}_m$ are the electric and magnetic fields of the $m$-th mode of the resonator, and ${\bf E}_i$ and ${\bf H}_i$ are the electric and magnetic fields of the incident wave, respectively. 

In the group-theoretical description, the selection rule (\ref{eq:3}) can be presented in the following form \cite{Overvig_PhysRevB_2020}:
\begin{equation}
\Gamma_{\partial_i}= \Gamma_{V}\otimes\Gamma_{m},
\label{eq:4}
\end{equation}
where $\otimes$ means the direct product of IRREPs,  $\Gamma_{\partial_i}$ defines the IRREP of a partial derivative of the Maxwell equations in the $i$ direction, $\Gamma_{V}$ is an IRREP of the perturbation operator, and $\Gamma_{m}$ is an IRREP of the unperturbed $m$-th mode (all the details of this approach can be found in reference \cite{Overvig_PhysRevB_2020}).

Equation (\ref{eq:4}) is the symmetry condition of nonzero interaction of the incident field and the $m$-th mode of the resonant system. From this equation, two conclusions follow. First, to be coupled with the magnetic field of the incident wave, the net magnetic dipole moment must be present in the field of the corresponding mode of the trimer. Such a moment does not exist in the eigenfield of the trimer belonging to the $C_{3v}$ group. In terms of electromagnetic theory, in this case, the overlap integral (\ref{eq:3}) of the incident plane wave and the toroidal dipole is equal to zero. However, the net magnetic moment can appear in the eigenfield of the trimer belonging to the $C_s$ group. Second, the magnetic field of the incident wave must have a component coinciding with the orientation of the net magnetic dipole moment of the eigenmode. We can confirm again that the reduction of the symmetry from $C_{3v}$ to $C_3$ does not provide the necessary existence of the net magnetic moment in the trimer (see table B2 in \ref{sec:appendixB}). 

Symmetry of the magnetic field $\bf H$ of the incident wave in the $x$-$y$ plane is described by the $C_{s}$ group. The $H_x$ and $H_y$ components of the magnetic field and the $m_x$ component of the magnetic dipole moment belong to IRREP $B_1$  of the $C_{s}$ group. The $m_y$ components of the magnetic dipole moment belongs to IRREP $A_1$. It means that if the incident field has only the $H_y$ component, the polar toroidal dipole moment cannot be excited in the trimer. Nevertheless, access to the toroidal dipole mode can be provided by the $H_x$ component of the magnetic field, because $H_x$ and $m_x$ have the same parity, i.e., they belong to the same IRREP $B_1$.

In the $C_{3v}$ symmetry, the $H_y$ component of the incident magnetic field excites the $m_y$ components in the disks forming the trimer [figure \ref{fig:fig5}(a)]. The projections of the magnetic dipole moments on the circle passing through the centers of the disks, cancel each other, and the resulting toroidal moment $\bf T$ is zero. The same is true for coupling with the incident wave having the $H_x$ component [figure \ref{fig:fig5}(b)]. In these two configurations, all disks forming the trimer resonate at the same frequency ($\omega_1=\omega_2=\omega_3$) [figure \ref{fig:fig5}(c)].

Further, we assume that a particular resonator in the trimer is perturbed. In this case, the resonant frequency of the perturbed resonator is shifted concerning the resonant frequency of the remaining two resonators in the trimer ($\omega_1\ne\omega_2=\omega_3$). Here, for the excitation of the toroidal mode, the orientation of the magnetic dipole moment in the perturbed resonator is important. Thus, when the trimer is excited by a wave having the $H_y$ component, the ${\bf m}_1$ component of the magnetic dipole moment in the perturbed resonator is orthogonal to the circle, the ${\bf m}_2$ and ${\bf m}_3$ components compensate each other, and the toroidal dipole mode does not arise [figure \ref{fig:fig5}(d)]. Contrariwise, if the incident wave has the $H_x$ component, the ${\bf m}_1$ component is tangent to the circle and the projections of the ${\bf m}_1$, ${\bf m}_2$, and ${\bf m}_3$ moments on the circle are no longer compensated [figure \ref{fig:fig5}(c)]. In this case, the net magnetic dipole moment appears, and the toroidal dipole moment arises at a frequency $\omega_\textrm{T}$ lying between the resonant frequencies of the individual disks [figure \ref{fig:fig5}(f)]. In this figure, ${\cal X}_1$, ${\cal X}_2= {\cal X}_3$ are the  imaginary parts of the impedances of the equivalent circuits  of dielectric  disks (see details in \ref{sec:appendixD}).
\subsection{Frequency and quality factor of toroidal dipole resonance}
Changing parameters of a particular disk in the trimer, e.g., its diameter $D$, thickness $h_d$, or refractive index $n$, one can realize access to the toroidal dipole mode of the trimer. However, such perturbations also affect on the  frequency  of the corresponding resonance. This influence can be evaluated qualitatively using equation (\ref{eq:2}) and the circuit theory. 

In the framework of the circuit theory, the trimer can be considered as a sequence of three ${\cal RLC}$ circuits. In the case of a small perturbation introduced into one resonator in the trimer, the resonant frequency of the mode can be obtained using equation (\ref{eq:124}) given in \ref{sec:appendixD}. In particular, for the diameter perturbation, the Taylor expansion of equation (\ref{eq:2}) in linear approximation and equation (\ref{eq:124}) yield the following  frequency shift $\Delta \omega$ of the toroidal mode:
\begin{equation}
\frac {\Delta \omega }{\omega_0}\approx-\frac{1}{3}\frac {\Delta D}{D},
\label{eq:circuit}
\end{equation}
where $\omega_0$ is the resonant frequency of the unperturbed trimer and $\Delta D$ is the value of the diameter perturbation. The coefficient ${1}/{3}$ is due to the perturbation of only one of the three resonators forming the trimer (this is consistent with equation (\ref{eq:127}) of the circuit theory given in \ref{sec:appendixD}). A similar expression can be derived for the case of the trimer perturbation via the refractive index of a particular disk.

The disk thickness does not enter to equation (\ref{eq:2}) as a parameter. To qualitatively evaluate the influence of the thickness perturbation on the toroidal mode resonance, one can consider that the resonant frequency of the disk is inversely proportional to its volume and, consequently, to its thickness.

The quality factor of a sharp Fano resonance is inversely proportional to the square of the coupling strength $\kappa$ \cite{Overvig_PhysRevB_2020}:
\begin{equation}
Q ={\cal P}/\kappa^2,
\label{eq:5}
\end{equation}
where ${\cal P}$ is the coefficient which depends on physical and geometrical parameters of the structure.  

\subsection{Metasurface with trimers in square unit cells}
In our foregoing analysis, we have found out that the toroidal mode can be excited in the metasurface composed of trimers with reduced symmetry by a normally incident plane wave with a proper polarization. In this section, we consider a metasurface composed of trimers arranged in square unit cells. Such a metasurface has  been investigated experimentally in \cite{Xu_AdvOptMater_2019}, but we present the main results here to ensure completeness of our study. Then we will modify the unit cell and reveal characteristics of a twin-trimer metasurface.

\begin{figure*}[!ht]
\centering
\includegraphics[width=0.65\linewidth]{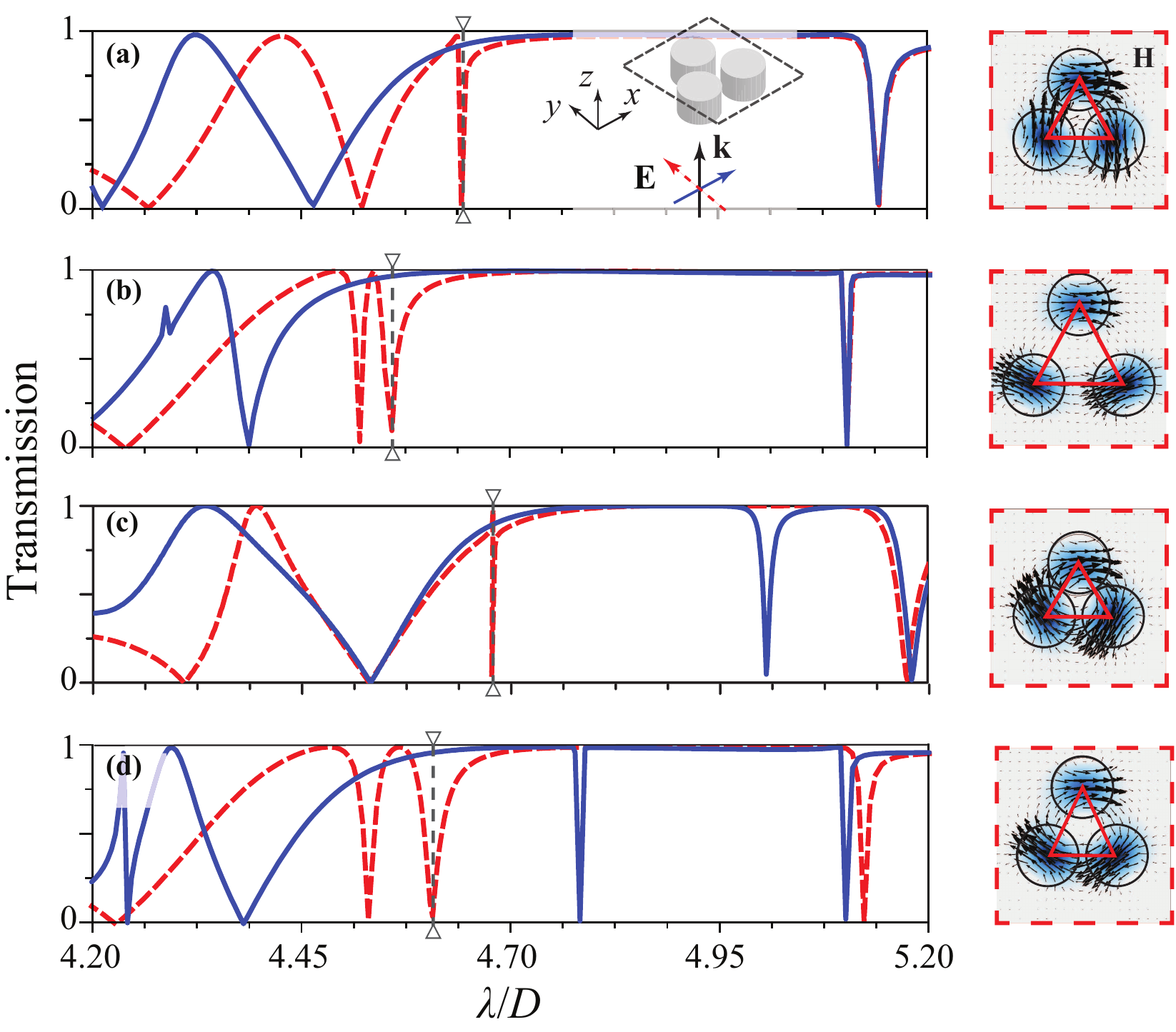}
\caption{Evolution of the transmission spectra versus normalized wavelength for the metasurface composed of square ($p\times p$) unit cells with (a), (b) non-perturbed trimers and (c), (d) trimers perturbed by a particular disk dislocation at the distance $\Delta h_t$ along the $y$ axis. Blue solid and red dashed lines correspond to the metasurface irradiation by the $x$- and $y$-polarized wave, respectively. In the schematics of the unit cell, the color map and black arrows correspond to the distribution of the magnetic field at the resonant wavelength. Modified parameters are: (a) $a_t/D=9/8$, $\Delta h_t=0$, (b) $a_t/D=12/8$, $\Delta h_t=0$, (c) $a_t/D=9/8$, $\Delta h_t=- 1/8$, and (d) $a_t/D=9/8$, $\Delta h_t=1/8$. All other parameters of the disks are the same as in figures \ref{fig:fig2} and \ref{fig:fig3} and the period-to-diameter ratio is $p/D=23/8$.}
\label{fig:fig6}
\end{figure*}

To simulate the eigenfields and optical response of the metasurface irradiated by a linearly polarized plane wave, we use the RF module of the commercial COMSOL Multiphysics finite-element electromagnetic solver. In the solver, we impose the Floquet-periodic boundary conditions on four sides of the unit cell to simulate an infinite two-dimensional array of the trimers. The solver allows us to calculate the transmission spectra of the metasurface as well as the resonant distribution of the inner electromagnetic field within the unit cell. 

We start our study from the eigenwaves analysis to recognize the appearance of the toroidal dipole mode of interest. From this analysis, the corresponding wavelength range is determined, where the toroidal mode exists in the given metasurface. Then in the chosen wavelength range, the transmission spectra of the metasurface is calculated for the incident waves of two orthogonal polarizations. 

The results of our simulation  of the metasurface composed of trimers disposed in square unit cells are collected in figure~\ref{fig:fig6}. This figure presents the transmission spectra evolution related to the toroidal dipole mode for non-perturbed and perturbed trimers with the $C_{3v}$ symmetry and its reduction, respectively, while keeping the symmetry of the square unit cell unchanged [see figure \ref{fig:fig4}(b)]. The resonance corresponding to the excitation of the toroidal mode appears only in the spectra of the $y$-polarized wave. We mark the position of this resonance on the wavelength scale by the grey arrows. The toroidal mode appears in the transmission spectra as a sharp resonance acquiring a Fano profile, where the dip in curve corresponds to the maximum of reflection, and the peak corresponds to the maximum of transmission. These extremes approach zero and unity, respectively, since the material losses are excluded in this simulation.

In particular, figure \ref{fig:fig6}(a) shows the transmission spectra of the metasurface with meta-molecules whose parameters coincide with those calculated above for a single trimer (see figure \ref{fig:fig3}). Comparing the results presented in these figures, one can conclude that the toroidal mode resonant wavelength in the metasurface spectra corresponds to the wavelength where the maximal contribution of the toroidal dipole mode to the scattering cross-section of the single trimer occurs. At the resonant wavelength, a particular near-field pattern appears within the unit cell, where the arrows of the magnetic field $\bf H$ demonstrate an in-plane circular flow which penetrates all three disks forming the cluster. Such a flow produces toroidal dipole moments oriented out-of-plane of the metasurface as predicted above. 

It is obvious that the toroidal mode is a hybrid mode of the trimer. The characteristics of the corresponding resonance depend on the degree of electromagnetic coupling between this hybrid mode and incident field. Therefore, when the distance between the disks increases, the area occupied by the trimer becomes larger. In this case, the electromagnetic coupling strength between the trimer and the incident field increases, which leads to a decrease in the quality factor of the corresponding resonance in the metasurface spectra as equation  (\ref{eq:5}) suggests. From figure \ref{fig:fig6}(b) one can conclude that the resonant wavelength acquires a blue shift for a larger trimer, while the in-plane circular flow of the magnetic field in the trimer preserves.
 
We also consider here the situation when the asymmetry of the trimer is enhanced by the dislocation of one of the resonators along the symmetry plane $\sigma$ [see figure \ref{fig:fig4}(c)]. The moving disk closer [figure \ref{fig:fig6}(c)] and farther [figure \ref{fig:fig6}(d)] leads to an increase and decrease in the quality factor of the corresponding resonance, respectively. This change in the quality factor is accompanied by a corresponding red and blue resonant frequency shift. Nevertheless, with the disk moving aside, the field distribution inside the trimer undergoes some distortion, where the magnetic field concentration in the dislocated disk decreases.

Regarding perturbations related to a change of the diameter, thickness, or material of one disk in the trimer [figure \ref{fig:fig4}(d)], we discuss them further when performing the symmetry reduction of twin-trimers disposed in the rhomboid unit cells.

Thus, if the trimer is located in a square unit cell, an electromagnetic coupling arises between the mode and the field of the incident wave. However, the ability to maintain high-$Q$ factor resonant conditions in such a structure is somewhat limited. The quality factor of the toroidal mode resonance increases as the disks in the trimer approach each other. However, in close proximity, the disks should not touch each other to ensure that the conditions for the existence of a hybrid mode of the trimer persist. To avoid this limitation, in what follows, we consider twin-trimer designs of the unit cell which allow to obtain a very high-$Q$ factor toroidal resonant response.

\section{Pairing toroidal dipole modes in twin-trimers}

\subsection{Reducing twin-trimer symmetry} 

\begin{figure}[t!]
\centering
\includegraphics[scale=0.3]{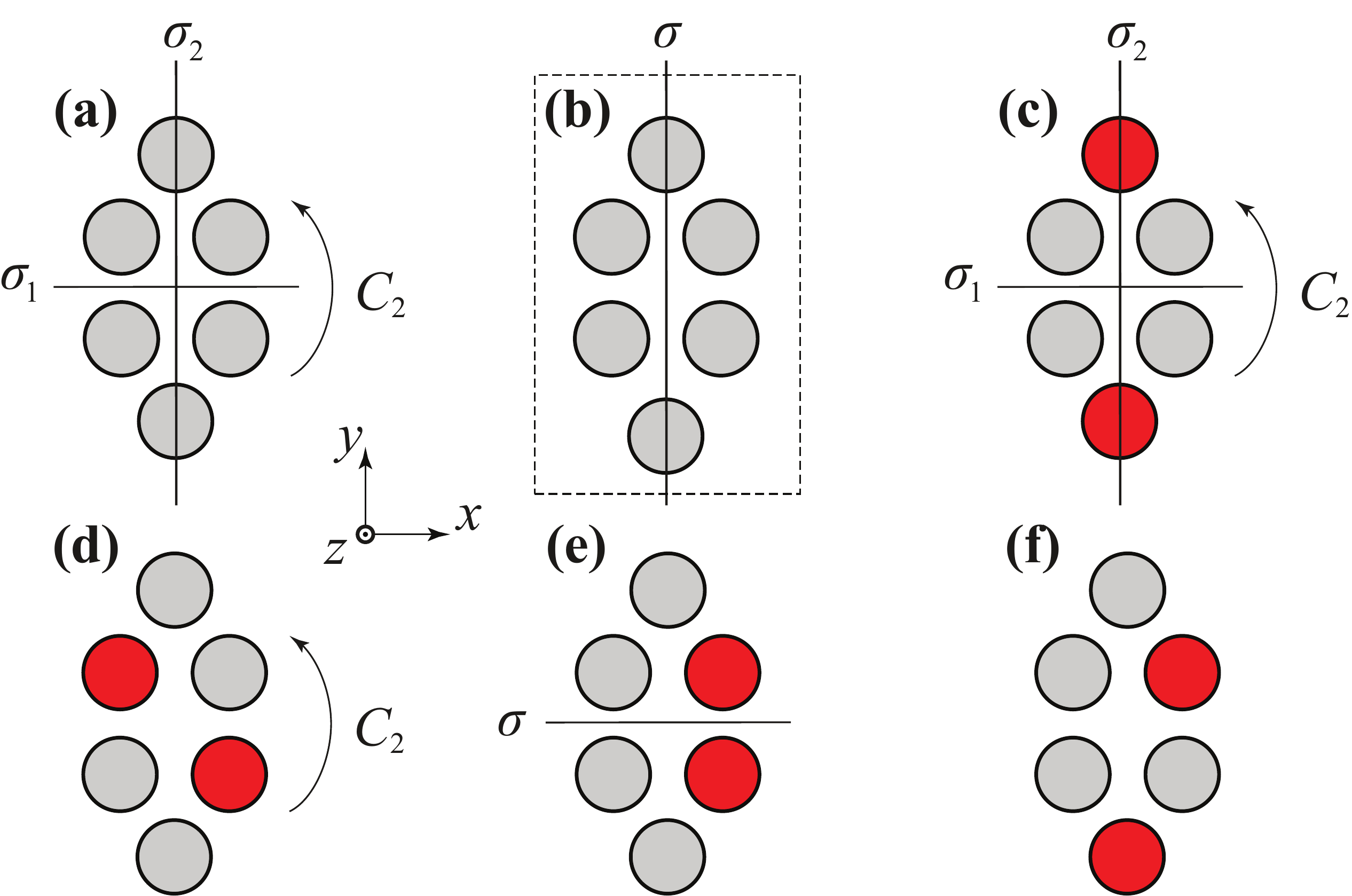}
\caption{Possible ways to reduce the twin-trimer symmetry, where (a) elements of the $C_{2v}$ symmetry of an unperturbed twin-trimer and perturbation to (b) the $C_{s}$ symmetry by dislocation of one of the disks, and  perturbations to (c) the $C_{2v}$, (d) $C_{2}$, (e) $C_{s}$, and (f) $C_{1}$ symmetry by substitution of two disks with other material or geometrical parameters.}
\label{fig:fig7}
\end{figure}

Several trimers can be combined together to form complex clusters (meta-macromolecules). In this section, we consider a metasurface whose meta-macromolecules are composed of two trimers disposed into either rectangular or rhomboid unit cells. In such clusters, there are two equal trimers, where one trimer is rotated by $180^\circ$ with respect to another trimer, as shown in  figure~\ref{fig:fig7}. Without any perturbation, i.e., with all six disks equal, the point symmetry of the twin-trimer corresponds to the $C_{2v}$ group. The symmetry elements of this group (which are two planes of symmetry $\sigma_1$ and $\sigma_2$ and one axis of the second order $C_2$) are shown in figure \ref{fig:fig7}(a). 

The twin-trimer deposition into a rectangular unit cell does not break the $C_{2v}$ symmetry. To make a perturbation, one disk can be dislocated along the $\sigma_2$ plane. This reduces the twin-trimer symmetry to the $C_{s}$ group [figure \ref{fig:fig7}(b)].

If one disk in each trimer forming the cluster is substituted by a disk with another material or with different geometrical parameters (these disks are indicated in figure \ref{fig:fig7} by red color), the $C_{2v}$ symmetry of the twin-trimer can be either preserved [figure \ref{fig:fig7}(c)], reduced to the $C_{2}$ group [figure \ref{fig:fig7}(d)], or reduced to the $C_{s}$ group [figure \ref{fig:fig7}(e)].  The twin-trimer without any symmetry can be also obtained [figure \ref{fig:fig7}(f)]. 

\subsection{Metasurface with twin-trimers in rectangular unit cells} 

\begin{figure*}[t!]
\centering
\includegraphics[width=1.0\linewidth]{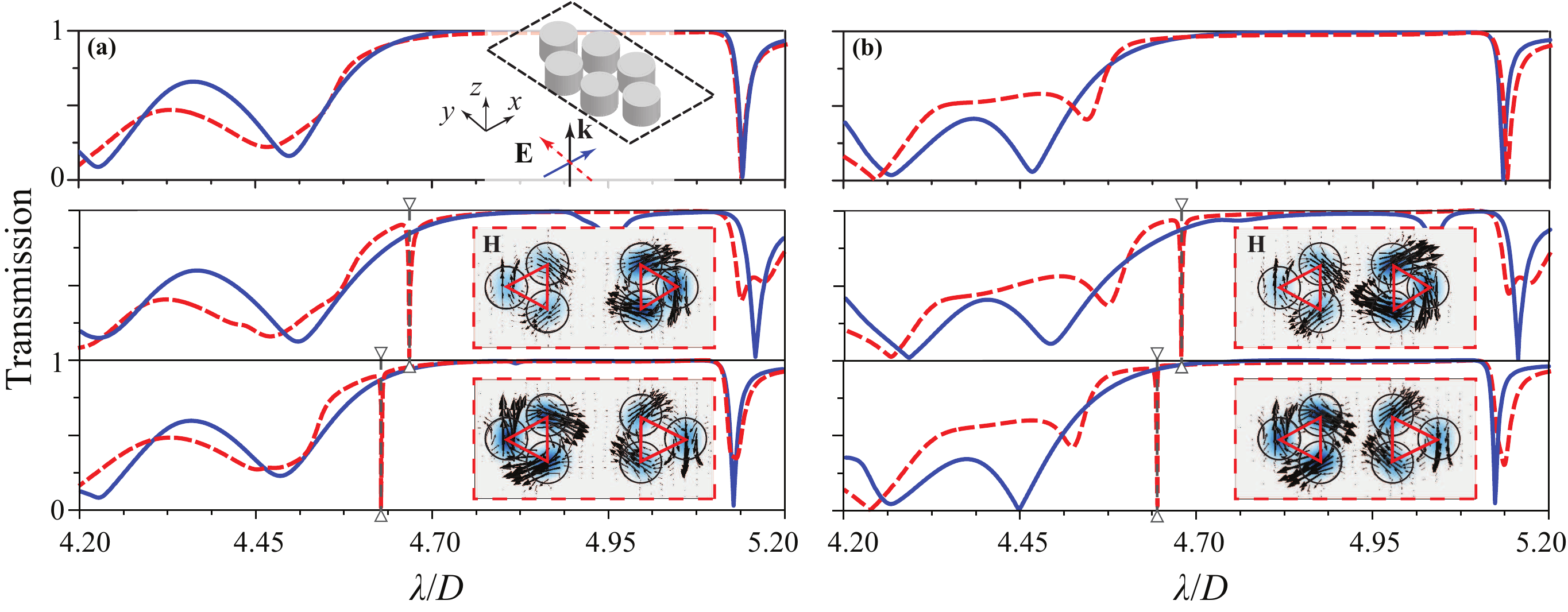}
\caption{The same as in figure \ref{fig:fig6} but for the metasurface composed of rectangular ($p\times 2p$) unit cells with twin-trimers. The distance between the trimers inside the unit cell is (a) $p$ and (b) $p-\Delta p$, where $p/D=23/8$, $\Delta p/D=4/8$, $a_t/D=9/8$, and $\Delta h_t=\pm 1/8$. In the insets, the magnetic field distribution at the corresponding resonant frequencies is shown for the perturbed twin-trimers.}
\label{fig:fig8}
\end{figure*}

Since twin-trimers in rectangular unit cells possess the $C_{2v}$ symmetry, the toroidal dipole mode turns out to be a quasi-dark state for such systems. The eigenwaves analysis shows that although the toroidal dipole mode exists in the twin-trimer cluster, it has a purely real eigenfrequency, and, therefore, has no electromagnetic coupling with the incident wave field. Breaking the symmetry of the twin-trimer by dislocating one disk  opens access to this state. This mechanism of excitation of the toroidal dipole mode in the twin-trimer metasurface is illustrated in figure \ref{fig:fig8}.

From this figure, one can conclude that as soon as the symmetry of the twin-trimer is broken, a resonance with high-$Q$ factor appears in the transmission spectra of the metasurface. This feature does not depend on which side the disk is dislocated [figure \ref{fig:fig8}(a)]. The quality factor of this resonance is the higher, the less the introduced perturbation. Thus it becomes possible to realize toroidal resonance with a very high-$Q$ factor, which is typical for resonances originating from the quasi-dark states \cite{Fedotov_PhysRevLett_2007}. 

As another perturbation parameter, we consider the distance between trimers inside the unit cell [figure \ref{fig:fig8}(b)]. Bringing the trimers closer slightly changes the transmission spectra of the metasurface, although it has little effect on the resonant frequency and quality factor of the toroidal dipole mode. The position of the resonance on the wavelength scale is practically independent of the distance between the trimers in the unit cell, which additionally confirms our conclusion that the properties of the toroidal dipole mode are determined primarily by the parameters of a single trimer and are not a consequence of the periodicity of the metasurface array.

At the same time, in the twin-trimer, there is an electromagnetic interaction between adjacent trimers, and the modes of individual trimers become to be coupled. As known, in the result of this interaction, the modes can be coupled either in-phase (symmetric) or out-of-phase
(anti-symmetric) reflecting correspondingly, the `bonding' and `anti-bonding' nature of such
configurations \cite{Rechberger_OptCommun_2003, Jain_ChemPhysLett_2010}. One can conclude that for the twin-trimer disposed in a rectangular unit cell, the net toroidal dipole mode acquires the bonding configuration. This configuration is characterized by in-phase rotation of the in-plane magnetic moments and, accordingly, by the co-directional arrangement of the resulting toroidal dipole moments in the two trimers forming the cluster. 

\subsection{Metasurface with twin-trimers in rhomboid unit cells} 
%

\begin{figure*}[!ht]
\centering
\includegraphics[width=0.65\linewidth]{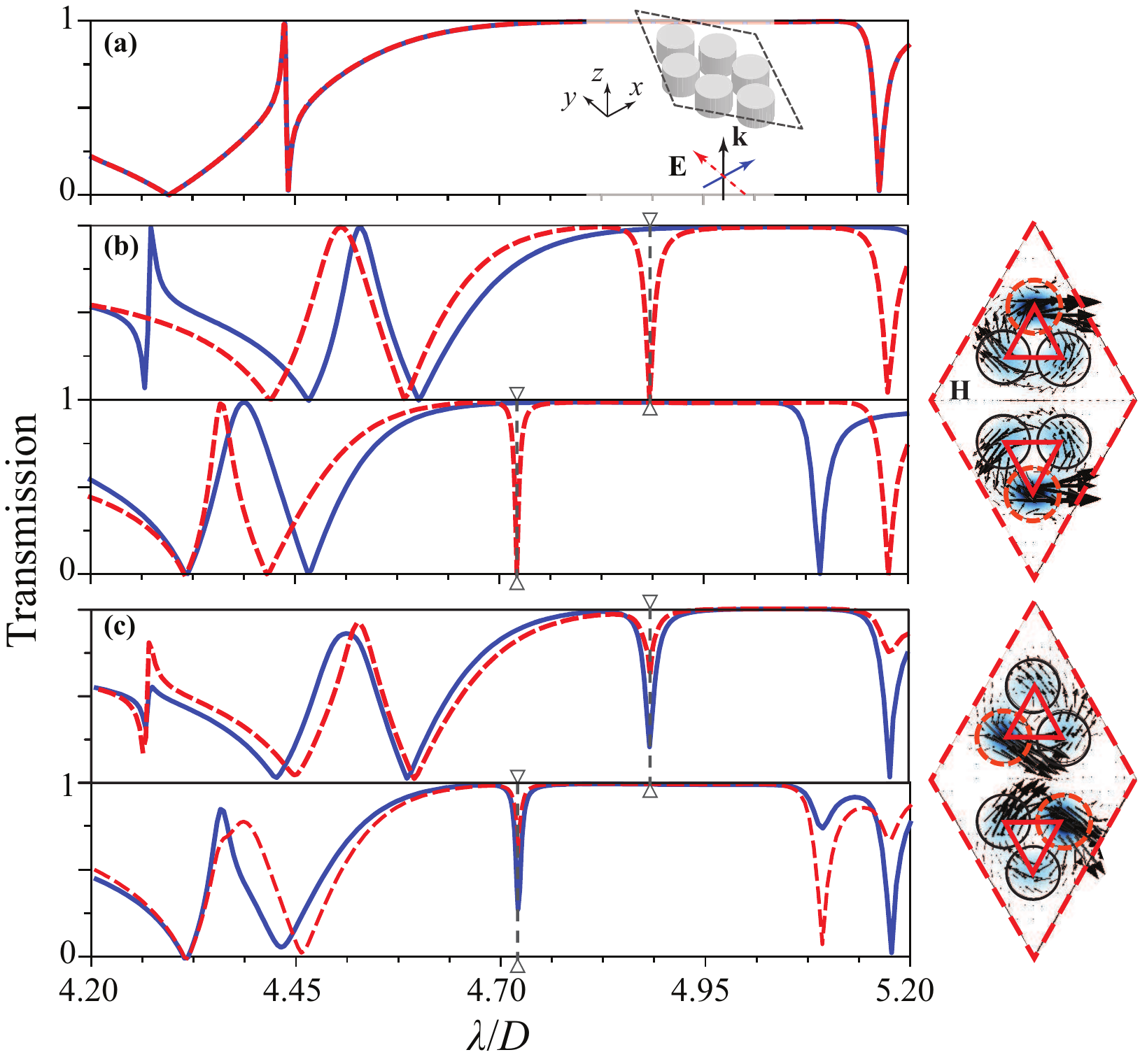}
\caption{The same as in figure \ref{fig:fig6} but for the metasurface composed of rhomboid unit cells with twin-trimers, where $p/D=27/8$. The perturbation is introduced into the thickness ($\Delta h_d=\pm 1/8$) of two disks in the cluster. In the insets, the resonant magnetic field distribution is presented for the case, when thickness of the perturbed disks increases (the perturbed disks are marked by red contours).}
\label{fig:fig9}
\end{figure*}

At the end of our theoretical consideration, in this section we study the manifestation of a toroidal dipole mode in metasurfaces composed of twin-trimers disposed in rhomboid unit cells. In particular, we present here the results of our numerical simulation for the metasurfaces, whose unit cells are disturbed by the out-of-plane symmetry breaking associated with the symmetries given in figures \ref{fig:fig7}(c) and \ref{fig:fig7}(d). As a perturbation parameter of the unit cell, we consider the change in the thickness of two disks in the twin-trimer. The results of our simulation of the transmission spectra of the metasurfaces and the corresponding inner resonant magnetic field distributions are summarized in figure \ref{fig:fig9}.

As reference data, figure~\ref{fig:fig9}(a) shows the characteristics of the metasurface composed of unperturbed twin-trimers. Unlike the structures considered so far, the response of this metasurface is the same on its irradiation by both $x$- and $y$-polarized waves. Therefore, such metasurface can be considered as a suitable platform for obtaining polarization-independent excitation of the toroidal dipole mode by a linearly polarized wave. However, as before, in the unperturbed structure, the toroidal dipole mode behaves as a dark state which does not appear in the metasurface spectra.

Figures~\ref{fig:fig9}(b) and \ref{fig:fig9}(c) present the change in the metasurface spectra with the twin-trimers perturbation. As soon as a perturbation is introduced into the unit cell, a resonance arises in the transmission spectra related to the manifestation of the toroidal dipole mode. By increasing the asymmetry parameter (thickness of two particular disks), the resonance shifts up, while decreasing that factor, the resonance shifts down on the wavelength scale. The quality factor of the resonance inversely depends on the degree of perturbation. It is noteworthy that for the structure characterized by the $C_2$ symmetry, the toroidal dipole mode resonance arises in the transmission spectra at the same wavelength for the waves of both polarizations.

The most distinctive feature of the metasurfaces with rhomboid unit cells is that the toroidal dipole mode coupling in the twin-trimers occurs in the anti-bonding fashion. This net toroidal dipole mode configuration is characterized by the out-of-phase distribution of the in-plane magnetic moments in two trimers, while the corresponding toroidal dipole moments appears to be oriented contra-directional. This effect can be considered as the manifestation of specific anti-toroidic order \cite{Spaldin_JPhys_2008}, which can be the subject of future research.

\section{Toroidal dipole mode manifestation in real metasurfaces} 
%

\begin{figure*}[!ht]
\centering
\includegraphics[width=0.5\linewidth]{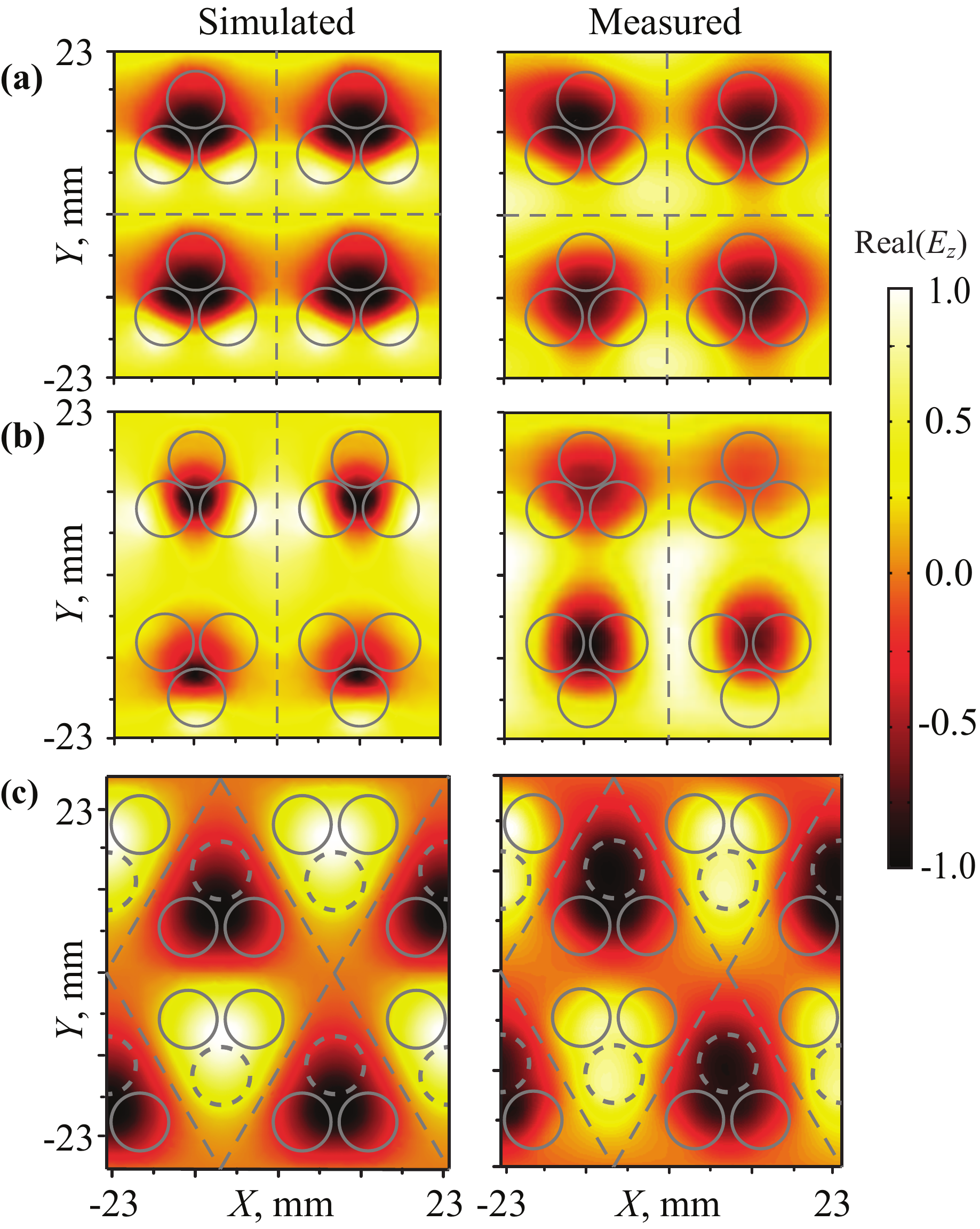}
\caption{Simulated and measured patterns of the real part of the $z$-component of the electric near-field for actual all-dielectric metasurfaces composed of (a) trimers in square unit cells, (b) twin-trimers in rectangular unit cells, and (c) twin-trimers in rhomboid unit cells. These patterns are plotted at the corresponding resonant wavelengths of the toroidal dipole mode excitation by the field of the $y$-polarized wave. Parameters of the samples are: $\varepsilon_d=23$, $\varepsilon_s=1.1$, $r_d=4$, $h_s=10$, $a_t=9$, (a) $p=23$, $h_d=5$, (b) $p=23$, $h_d=5$, $\Delta h_t=-1$, $\Delta p=-1$, and (c) $p=27$, $h_d=3.5$, and $\Delta h_d=1$. All geometrical parameters are given in millimeters. The measurements are performed at the distance of 1~mm above the metasurface.}
\label{fig:fig10}
\end{figure*}

In this section, the goal is to confirm our theoretical findings and find out the resistance of the toroidal mode to internal losses inherent in real metasurfaces. To this end, we apply a microwave approach \cite{Xu_AdvOptMater_2019, Sayanskiy_PhysRevB_2019, Kupriianov_PIERS_2019} for prototyping and investigation resonant features of all-dielectric metasurfaces under consideration. 

In our experimental study, we utilize an inexpensive technology, where particles are mechanically fabricated from commercially available ceramic plates. The particles are assembled into metasurfaces whose microwave characterization is performed by using standard equipment. This approach allows us to measure the distribution of both magnetic and electric components of the near-field in close proximity above the metasurface. In combination with the far-field analysis, it gives us a complete picture of the appearance of the toroidal dipole mode in real metasurfaces.

In our approach, the microwave ceramic is used as a dielectric material for the disks fabrication. According to the manufacturer's data sheet, the relative permittivity of this ceramic is $22 \pm 1$ and loss tangent is $\tan \delta \le 1 \times 10^{-3}$ referenced at $10$~GHz. The disks are fabricated with the use of a precise milling machine. The production resolution of the machine is $0.1$~mm. The diameter of the disks is $8$~mm. 

In our study, we utilize two sets of disks with thicknesses $3.5$~mm and $5$~mm. An additional set of disks with thickness of $2.5$~mm is prepared for their use to compose `perturbed' trimers. To arrange the disks into a lattice, special arrays of holes are milled in holders made of rigid-foam plates. The depth of holes in the holders corresponds to the thickness of the disks used. 

At the first stage, we perform the far-field measurements to find out the corresponding resonances in the metasurface transmission spectra. For these measurements, we use a pair of antennas operating in the frequency range from $8.2$~GHz to $12.4$~GHz. The measurements are performed with the use of the Vector Network Analyzer (VNA) Rohde \& Schwartz ZVA50. All resonances of interest can be readily recognized in the transmission spectra, although they are somewhat suppressed by actual losses existing in the metasurfaces (corresponding spectra are not presented here, partially they can be found in \cite{Xu_AdvOptMater_2019, Kupriianov_PIERS_2019}).

After the position of the toroidal mode resonance is determined on the frequency scale, we perform the mapping of near-fields to plot the electric field distribution close to the metasurface plane. In particular, to measure the distribution of the real part of the $z$-component of the electric field, the sample is irradiated by a quasi-plane wave with a proper polarization, while the dipole probe is situated normally to the sample surface. The near-field imaging system is used to perform the movement of the probe along the metasurface plane. The scanning area covers several unit cells (see details on the measurement technique in references \cite{Xu_AdvOptMater_2019, Sayanskiy_PhysRevB_2019, Kupriianov_PIERS_2019}). The results of our measurements of the corresponding near-fields are presented in figure \ref{fig:fig10}. The obtained plots allow us to reveal all characteristics of the toroidal dipole mode, which appears in three different trimer-based metasurfaces. 

Foremost, one can conclude that our measured data are in reasonable agreement with the simulation and indeed demonstrate the behaviours typical for the toroidal dipole mode discussed above. In the first design the symmetry of the trimer is broken by the square unit cell [figure \ref{fig:fig10}(a)], while in the second design, one disk in the twin-trimer is slightly moved [figure \ref{fig:fig10}(b)]. For the first and second designs of the cluster, the toroidal moments in the neighboring trimers are oriented identically and directed out-of-plane. Therefore, the electric near-field is strongly localized in the centers of trimers forming hot-spot outside the dielectric particles. Such an appearance of the electric near-field obtained in the experiment is evidence of the toroidal dipole mode excitation.

In the third design, two trimers are paired together in such a way that a rhomboid cluster appears [figure \ref{fig:fig10}(c)]. To excite the toroidal moments by a normally incident plane wave, the out-of-plane symmetry breaking is introduced into the clusters. This symmetry breaking is realized by substituting disks with other thickness (the disks responsible for the symmetry breaking are depicted in figure \ref{fig:fig10}(c) by grey dashed contours). On can conclude that the toroidal dipole mode is excited in the cluster in the antiparallel fashion, where the electric near-field hot-spots have the staggered arrangement in the neighboring trimers.

\section{Concluding remarks}  
In this paper, we have suggested a systematic study of conditions for the appearance of a toroidal dipole mode in all-dielectric metasurfaces. Our theoretical analysis involves the techniques from the point group theory, SALC, and circuit theory. A cluster of three particles (trimer) is considered as a structural element (meta-molecule) of the metasurface. The developed approach is also applied to the metasurface composed of more complex twin-trimer clusters (meta-macromolecules). Our theoretical findings are supplemented by both the full-wave numerical simulation and experimental study.

We have shown that selection rules for the toroidal dipole mode can be expressed in the explicit electromagnetic formulation and described in the group-theoretical language. The latter provides a simple and efficient way to choose a corresponding design of the metasurface supporting toroidal dipole mode. We have shown how the quality factor of the toroidal resonance depends on the asymmetry introduced into the trimer-based unit cell.

Presented above results show that symmetry of the complex system is various for different complexity levels. In our case, the symmetry of the disk is $C_{\infty v}$, the symmetry of the magnetic dipole and perturbed trimer is $C_{s}$, and the symmetry of the perturbed twin-trimer can be either $C_{2v}$, $C_{2}$, or $C_{s}$. 
 
The structure of the inner electromagnetic field of the toroidal dipole mode is defined by the local symmetry of the trimer, which in the perturbed case is $C_{s}$.  However, the far-field characteristics depend on the global symmetry of the meta-(macro)molecule. In general, this problem can be associated with the theory of point defects in crystals.

We have noticed that the intensity of the electromagnetic field inside the perturbed resonator can be higher or lower than that in the remaining two resonators in the trimer. This perturbed resonator can be considered as an active center for other resonators forming the cluster.

Our results can provide remarkable opportunities for facilitating the interaction of light and matter in all-dielectric structures intended for light emission, nonlinear switching, sensing, and so on.

\section*{\label{ack}Acknowledgments}
VD thanks the Brazilian Agency National Council of Technological and Scientific Development (CNPq) for financial support. SDSS acknowledges support form CNPq (Grant No.~160344/2019-0). ASK and VRT acknowledge financial support from the National Key R\&D Program of China (Project No.~2018YFE0119900).

\section*{\label{orcid}ORCID iDs}
Victor Dmitriev\\ https://orcid.org/0000-0002-7715-7300\\
\\
Anton S Kupriianov\\ https://orcid.org/0000-0002-1048-442X\\
\\
Silvio Domingos Silva Santos\\
https://orcid.org/0000-0002-0557-2492\\
\\
Vladimir R Tuz\\ https://orcid.org/0000-0001-6096-7465

\appendix 
\section{Transformation properties of magnetic dipole  moment}
\label{sec:appendixA}
The transformation formula for the magnetic dipole vector under rotation-reflections has the following form:
\begin{equation}
 {\bf m}^{\prime} =\det (\!{\bf{\bar {\,D}}}) {\bf{\bar {\,D}}}\bf m,
\label{eq:A1}
\end{equation}
where $\bf m$ is the magnetic dipole moment, ${\bf m}^{\prime}$ is the mapped vector, $\bf{\bar {\,D}}$ is a 2D representation of the rotation-reflection symmetry element $R$, and $\det$ means determinant. In the case of trimers, one has $\det(\!{\bf{\bar {\,D}}})=1$ for the three-fold rotation $C_{3}$ around the $z$ axis, and $\det(\!{\bf{\bar {\,D}}})=-1$ for the reflections with respect to planes $\sigma_x$ and $\sigma_y$. The magnetic field $\bf H$ is an axial vector, therefore, it is transformed in accordance with equation (\ref{eq:A1}) as well. 

\section{Tables of group theory}
\label{sec:appendixB}
%

\begin{table*}[t!]
\begin{center}
\caption{IRREPs of the $C_{3v}$ group and SALCs of unperturbed trimer in ${\bf m}_i$ basis}
{
\begin{tabular}{c@{\hspace{0mm}}c@{\hspace{0mm}}c@{\hspace{0mm}}
c@{\hspace{0mm}}c@{\hspace{0mm}}c@{\hspace{0mm}}c@{\hspace{0mm}}
c@{\hspace{0mm}}} \hline
\fbox{$C_{3v}$} & e & $C_3$ & $C_3^{-1}$ & $\sigma_1$ & $\sigma_2$ & $\sigma_3$ & SALCs \\  
\hline
\\
$A$ & 1 & 1 & 1 & 1 & 1 & 1 & $\frac{{\bf m}_2+{\bf m}_4+{\bf m}_6}{\sqrt{3}}$ \\ 
\\
$B$ & 1 & 1 & 1 & $\!\!\!\!\! -1$ & $\!\!\!\!\! -1$
& $\!\!\!\!\! -1$ & $\frac{{\bf m}_1+{\bf m}_3+{\bf m}_5)}{\sqrt{3}}$  
\\\\
$E$ &
$\Biggl(\!\!\!
\begin{array} {cc}
    1 & 0   \\
    0 & 1
\end{array} \!\!\!
\Biggr)$ &
$\left(\!\!\!
\begin{array} {cc}
    \!\!\!-\frac{1}{2} & \!\!\!-\frac{\sqrt{3}}{2}   \\
    \frac{\sqrt{3}}{2} & -\frac{1}{2}
\end{array} \!\!\!
\right)$ &
$\left(\!\!\!
\begin{array} {cc}
    \!\!\!-\frac{1}{2} & \frac{\sqrt{3}}{2}   \\
    -\frac{\sqrt{3}}{2} & -\frac{1}{2}
\end{array} \!\!\!
\right)$ &
$\Biggl(\!
\begin{array} {cc}
    \!\!\!\!-1 & 0   \\
    0 & 1
\end{array}\!\!\!
\Biggr)$ &
$\left(\!\!\!
\begin{array} {cc}
    \frac{1}{2} & \frac{\sqrt{3}}{2}   \\
    \frac{\sqrt{3}}{2} & \!\!\!-\frac{1}{2}
\end{array} \!\!\!
\right)$ &
$\left(\!\!\!
\begin{array} {cc}
    \frac{1}{2} & -\frac{\sqrt{3}}{2}   \\
    -\frac{\sqrt{3}}{2} & \!\!\!-\frac{1}{2}
\end{array} \!\!\!
\right)$  &
$\begin{array} {c} \frac{{2\bf m}_1-{\bf m}_3-{\bf m}_5}{\sqrt{6}}  \pm \frac{{\bf m}_4-{\bf m}_6}{\sqrt{2}} \\
\frac{{2\bf m}_2-{\bf m}_4-{\bf m}_6}{\sqrt{6}} \pm \frac{{\bf m}_3-{\bf m}_5}{\sqrt{2}}
\end{array}$ \\
\\
\hline		
\end{tabular}
}
\end{center}
\label{tab:tab1}
\end{table*}

\begin{table*}[h!]
\begin{center}
\caption{IRREPs of the $C_{s}$ and $C_{3}$ groups and symmetry degeneration of the $C_{3v}$ group to the $C_s$ and $C_3$ subgroups} 
{
\begin{tabular}{c@{\hspace{4mm}}c@{\hspace{4mm}}c@{\hspace{4mm}}
} 
\hline
IRREPs of the $C_s$ group and transformation \\ properties of magnetic field ${\bf H}$ and magnetic & IRREPs of the $C_3$ group, & Symmetry \\ moment ${\bf m}$ in perturbed trimer in the $x$-$y$ basis & $\varepsilon=e^{\rmi 2\pi/3}$ & degeneration \cite{Tasolamprou_PhysRevB_2016} \\  
{
\begin{tabular}{c@{\hspace{4.0mm}}c@{\hspace{4.0mm}}c@{\hspace{4.0mm}}c}
\hline
\fbox{$C_{s}$} & $e$ & $\sigma_1$ & Magnetic dipole moment, \\
			   &     & ($x=0$)      &   magnetic field \\
\hline
$A_1$ & 1 & 1 & $m_y$ \\
$B_1$ & 1 & $\!\!\!\!-1$ & $m_x$, $H_x$, $H_y$ \\
\end{tabular}
}
    & \begin{tabular}{c@{\hspace{4.0mm}}c@{\hspace{4.0mm}}c@{\hspace{4.0mm}}c}
	\hline
	\fbox{$C_{3}$} & $e$ & $C_3$ & $C_3^{-1}$ \\
	\hline
	$A$ & 1 & 1 & 1 \\
	\raisebox{-1ex}{E} & 1 & $\varepsilon$ & $\varepsilon^{\star}$ \\
				       & 1 & $\varepsilon^{\star}$ & $\varepsilon$ \\
	\end{tabular}
 	& \begin{tabular}{c@{\hspace{4.0mm}}c@{\hspace{4.0mm}}c}
	\hline
	$C_{3v}$ & $C_{s}$  & $C_{3}$ \\
	\hline
	$A$ & $A_1$ & $A$ \\
	$B$ & $B_1$ & $A$ \\
	$E$ & $B_1$ & $E$ \\
	\end{tabular} \\
 	\hline
    \end{tabular}
}
\end{center}
\label{tab:tab2}
\end{table*}
%
IRREPs of the $C_{3v}$ group and SALCs of an unperturbed trimer are presented in table B1. Table B2 shows IRREPs of the $C_s$ group and transformation properties of the magnetic field ${\bf H}$ and magnetic moment ${\bf m}$ in a perturbed trimer, IRREPs of the $C_3$ group, and symmetry degeneration of the $C_{3v}$ group into the $C_s$ and $C_3$ groups.
\section{Symmetry-adapted linear combination (SALC) approach}
\label{sec:appendixC}
%
%
\begin{figure}[!ht]
\centering
\includegraphics[scale=0.3]{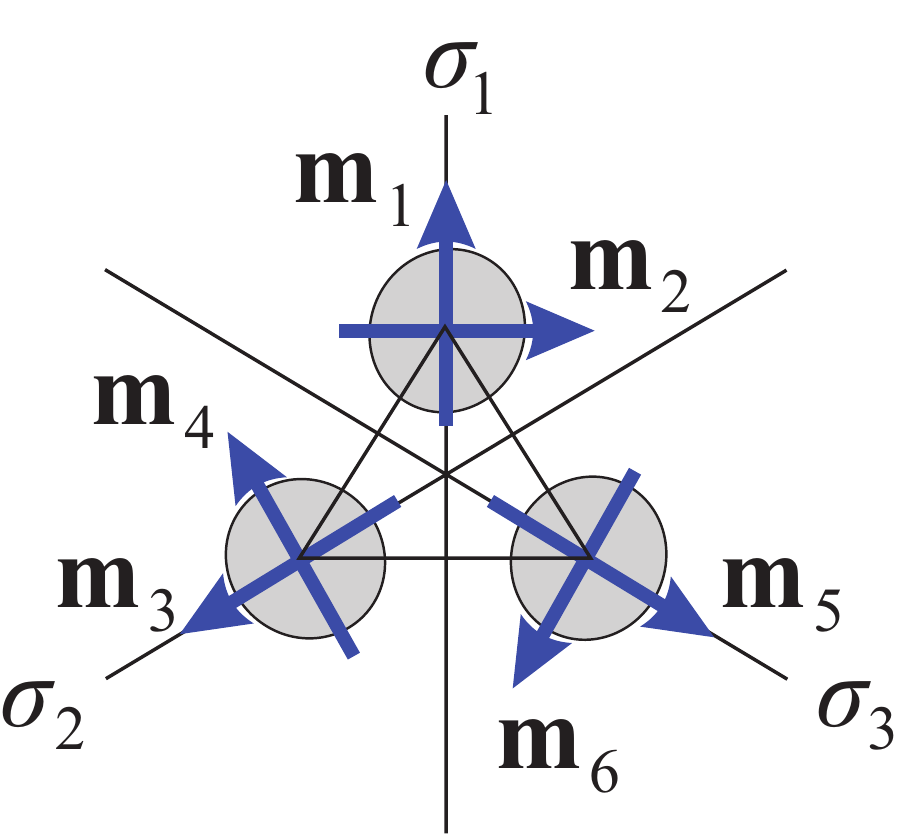}
\caption{Symmetry adapted basis for the trimer with the $C_{3v}$ symmetry.}
\label{fig:figC1}
\end{figure}

The chosen basis for dipole magnetic moments is shown in figure \ref{fig:figC1}. The unit vectors ${\bf m}_i$ with odd and even index $i$ are oriented parallel and perpendicular to the planes of symmetry, respectively. Details of this group-theoretical technique with application to a mechanical vibrating system can be found in \cite{McWeeny, Wolbarst}. One can calculate SALCs using the projection operators \cite{Bradley_book_2009}:
\begin{equation}
{\bf m}_{kj}^{\Gamma} \!=\! \dfrac{1}{g} \sum_{R}\chi_{kj}^{\Gamma}(R)R{\bf m}_i,
\label{eq:22}
\end{equation}
where the summation is performed with respect to the elements $R$ of the $C_{3v}$ group. These elements are $e$, $C_3$, $C_3^{-1}$, $\sigma_{1}$, $\sigma_{2}$, and $\sigma_{3}$. $\Gamma$ is a representation $A$, $B$, or $E$ of the $C_{3v}$ group (see table B1 in \ref{sec:appendixB}), $\chi_{kj}^{\Gamma}(R)$ is the $kj$-th component of the IRREP $\Gamma$ of the element $R$, $R{\bf m}_i$ is a basis element ${\bf m}_i$ transformed by the operator $R$, and $g$ is the order of the group (in our case, $g=6$). 

The normalized eigenvectors of the trimer are presented in the third column of table C1. Corresponding eigenmodes are also identified performing the full-wave simulation in reference \cite{Tasolamprou_PhysRevB_2016}. Notice that the dipole modes $D_x$ and $D_y$ are degenerate because they belong to the same 2D IRREP $E$ and are orthogonal being on different lines of this IRREP. The same is true for the quadrupole modes $Q_1$ and $Q_2$.
\begin{table*}[ht!]
\begin{center}
\caption{Eigenmodes of trimer with the $C_{3v}$ symmetry in terms of vectors ${\bf m}_i$}
{
\begin{tabular}{c@{\hspace{4mm}}c@{\hspace{4mm}}c@{\hspace{4mm}}
} \hline
\\
IRREP & Eigenmode image & Description \\  
\\  
\hline
\hline
\raisebox{7ex}{$A$} 
& \includegraphics[height=2.5cm]{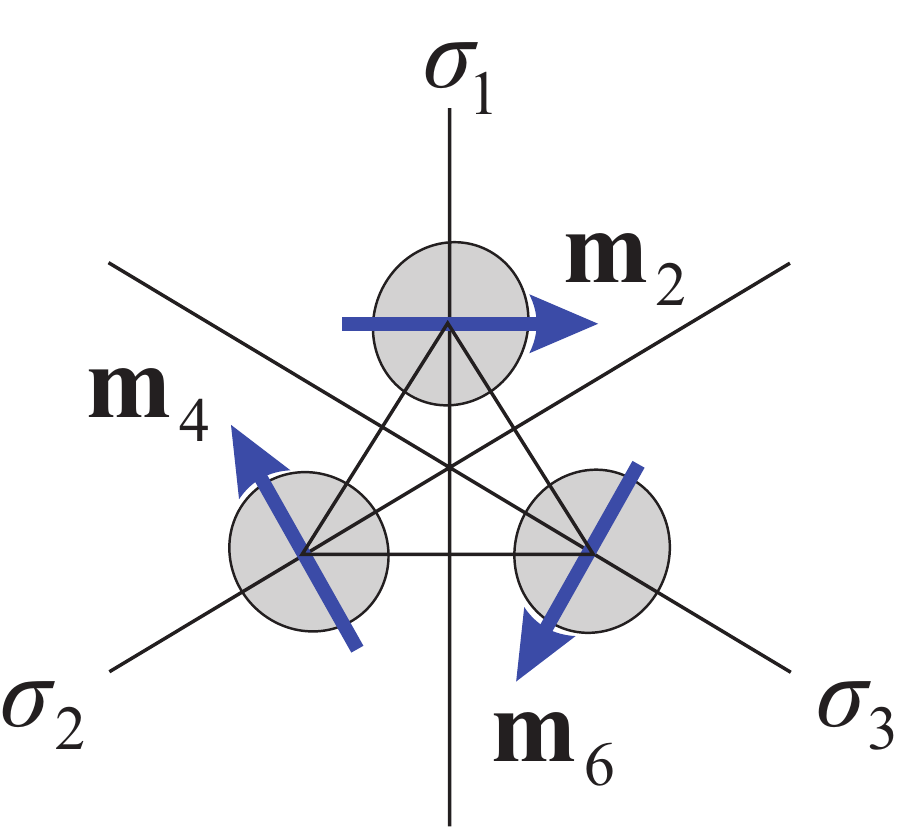}
& \raisebox{7ex}{
    $\begin{array}{c}
    \textrm{Toroidal mode} \\
    \dfrac{{\bf m}_2+{\bf m}_4+{\bf m}_6}{\sqrt{3}}
    \end{array}$
}
\\
\hline
\hline
\\
\raisebox{7ex}{$B$}
& \includegraphics[height=2.5cm]{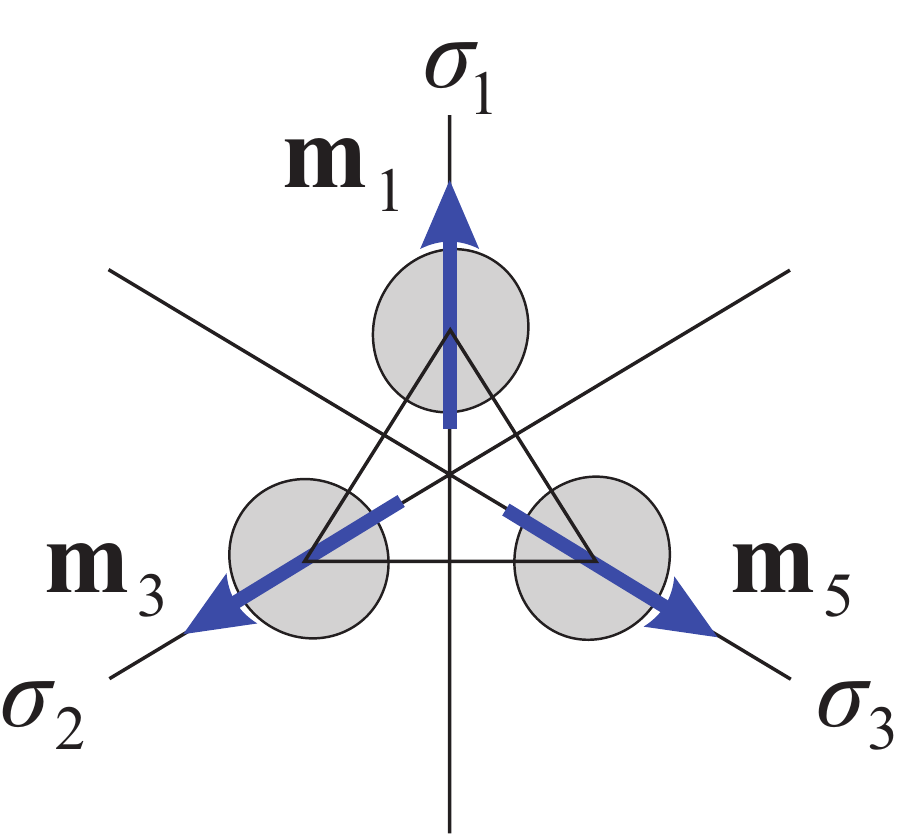}
& \raisebox{7ex}{
    $\begin{array}{c}
    \textrm{Radial mode} \\
    \dfrac{{\bf m}_1+{\bf m}_3+{\bf m}_5}{\sqrt{3}}
    \end{array}$
}
\\               
\hline
\hline
\raisebox{7ex}{
    $\begin{array}{c}
    E \\
    \textrm{first line}
    \end{array}$
}  
& \includegraphics[height=2.5cm]{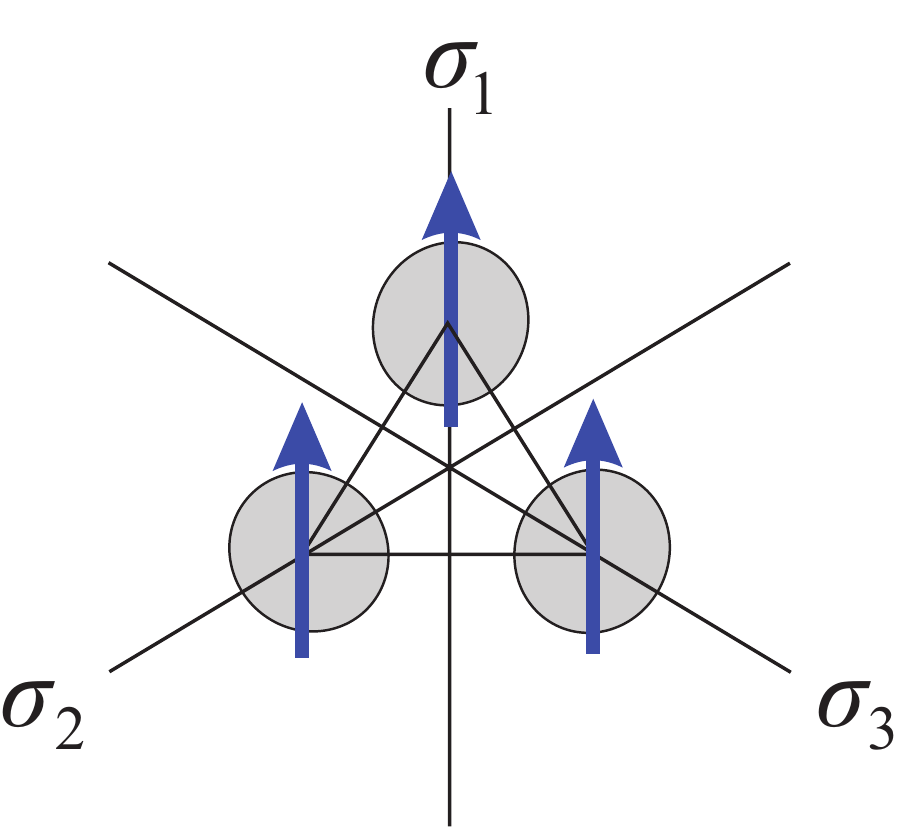}
& \raisebox{7ex}{
    $\begin{array}{c}
    \textrm{Dipole mode}~D_y \\
    \dfrac{{2\bf m}_1-{\bf m}_3-{\bf m}_5}{\sqrt{6}} + \dfrac{{\bf m}_4-{\bf m}_6}{\sqrt{2}}
    \end{array}$
}	
\\
\hline
\raisebox{7ex}{
    $\begin{array}{c}
    E \\
    \textrm{first line}
    \end{array}$
}
& \includegraphics[height=2.5cm]{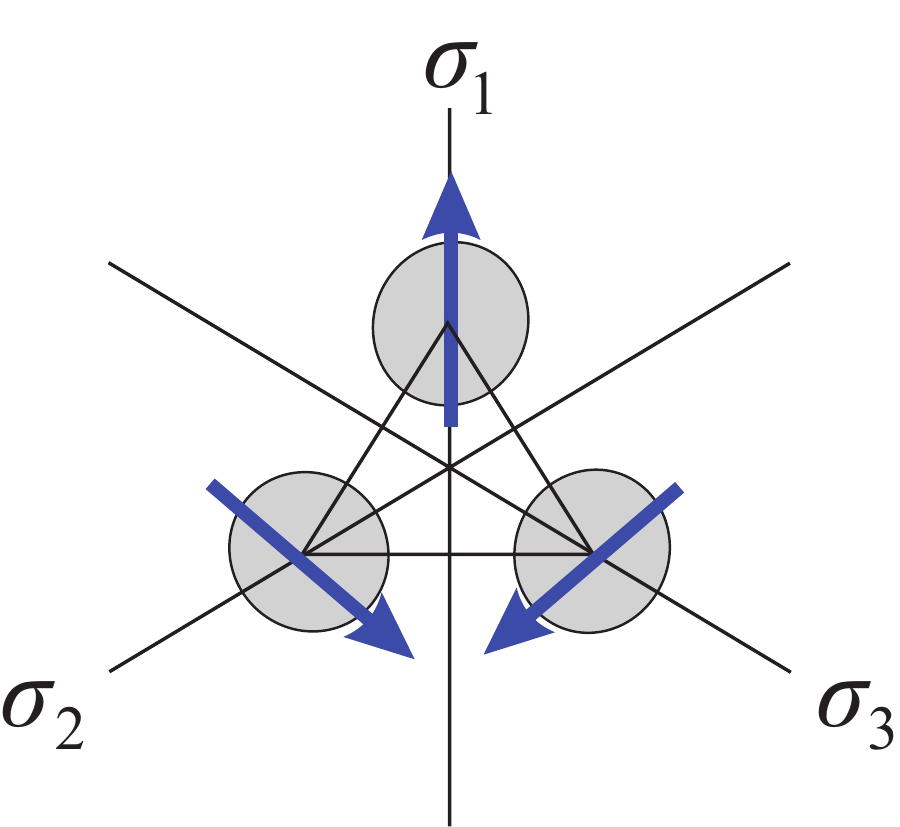}
& \raisebox{7ex}{
    $\begin{array} {c}
    \textrm{Quadrupole mode}~Q_1 \\
    \dfrac{{2\bf m}_1-{\bf m}_3-{\bf m}_5}{\sqrt{6}} - \dfrac{{\bf m}_4-{\bf m}_6}{\sqrt{2}}
    \end{array}$
}
\\
\hline
\raisebox{7ex}{
    $\begin{array}{c}
    E \\
    \textrm{second line}
    \end{array}$
}   
& \includegraphics[height=2.5cm]{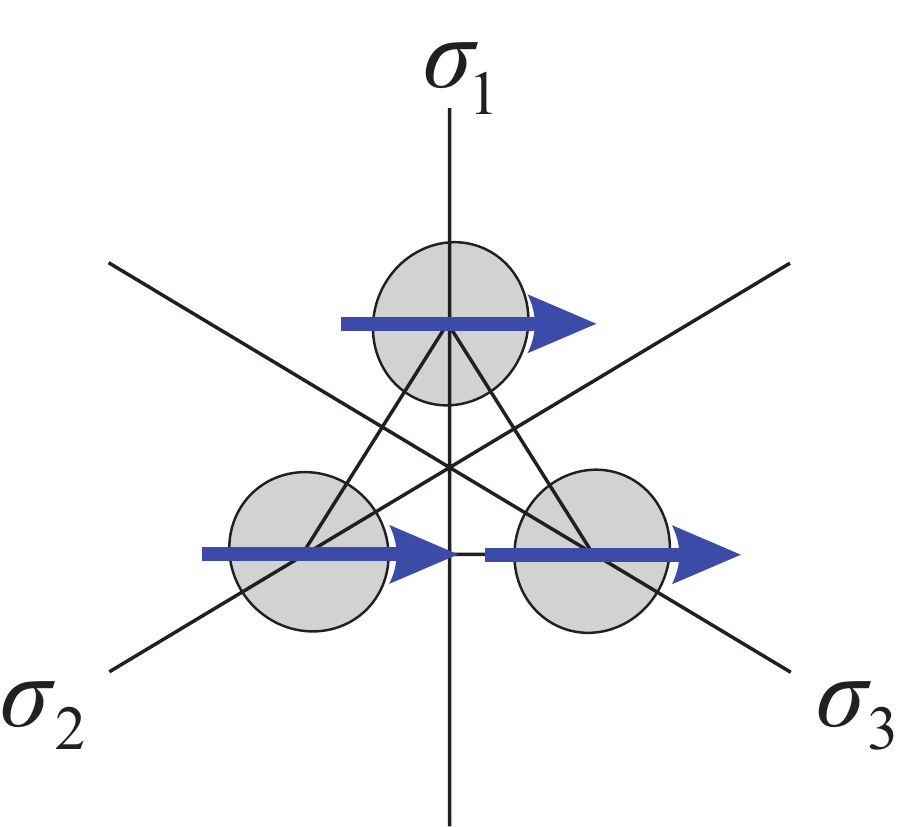}
& \raisebox{7ex}{
    $\begin{array}{c}
    \textrm{Dipole mode}~D_x \\
    \dfrac{{2\bf m}_2-{\bf m}_4-{\bf m}_6}{\sqrt{6}} - \dfrac{{\bf m}_3-{\bf m}_5}{\sqrt{2}}
    \end{array}$
}	 		   		
\\
\hline 		   		
\raisebox{7ex}{
    $\begin{array}{c}
    E \\
    \textrm{second line}
    \end{array}$
}
& \includegraphics[height=2.5cm]{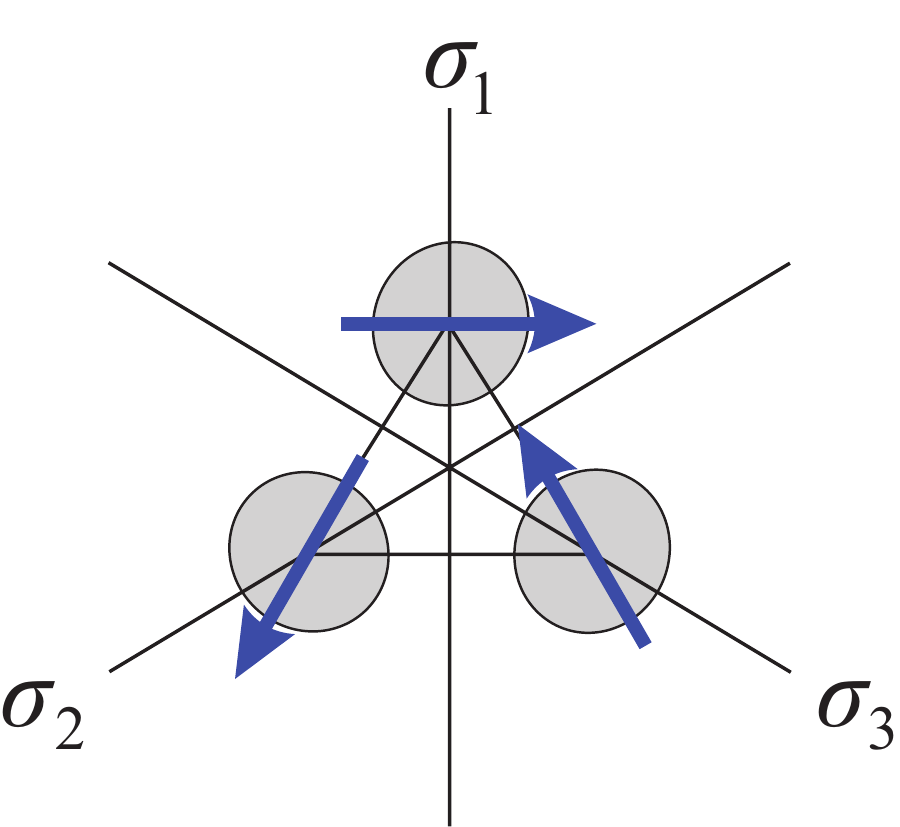}
& \raisebox{7ex}{
    $\begin{array}{c}
     \textrm{Quadrupole mode}~Q_2 \\
    \dfrac{{2\bf m}_2-{\bf m}_4-{\bf m}_6}{\sqrt{6}} + \dfrac{{\bf m}_3-{\bf m}_5}{\sqrt{2}}
    \end{array}$
}
\\
\hline	
\hline
\end{tabular}
}
\end{center}
\label{tab:tab3}
\end{table*}
%
\section{Circuit theory of trimer}
\label{sec:appendixD}
An equivalent ${\cal RLC}$ circuit of a dielectric disk at a resonance can be described via the impedance ${\cal Z}$=${\cal R}+\rm{j}{\cal X}$, where  ${\cal R}$ is the resistance and ${\cal X}$ is the reactance. If we consider a series ${\cal RLC}$ scheme, the impedance and resonant frequency are ${\cal Z}={\cal R}+\rm{j}(\omega {\cal L}-1/\omega {\cal C})$ and $\omega_0=1/\sqrt{\cal L\cal C}$, respectively. 

The reactance slope parameter $x$ of the circuit is defined as 
\begin{equation}
x=\dfrac{\omega_0}{2} \frac{d {\cal X}}{d\omega}{\biggl|}_{\omega=\omega_0}=\omega_0 {\cal L}=\frac{1}{\omega_0 {\cal C}}.
\label{eq:122}
\end{equation}
In a narrow-band case, the reactance is
\begin{equation}
{\cal X} \approx \omega_0 {\cal L} \dfrac{2\Delta\omega}{\omega_0}=x\dfrac{2\Delta\omega}{\omega_0},
\label{eq:123}
\end{equation}
where $\Delta\omega=(\omega-\omega_0)$. If $\cal Z$ includes internal and radiation losses, the loaded quality factor of the circuit is 
\begin{equation}
Q_L=\dfrac {\omega_0 {\cal L}}{\cal R} = \dfrac {x}{\cal R},
\label{eq:129}
\end{equation}
and the $-3$dB fractional bandwidth (BW) of the resonator is defined as $\textrm{BW}=\omega_0/Q_L$.

A trimer composed of three resonators can be represented by three $\cal R$$\cal L$$\cal C$ circuits connected in series. The resonant condition for this contour is 
\begin{equation}
{\cal X}_1+{\cal X}_2+{\cal X}_3=0.
\label{eq:124}
\end{equation}
For the perturbed trimer discussed above in the main body of this paper, we have ${\cal X}_3={\cal X}_2$ and $\omega_3=\omega_2$. From equations (\ref{eq:123}) and (\ref{eq:124}) one can derive the condition for the resonant frequency of the toroidal mode $\omega_\textrm{T}$ in the trimer:
\begin{equation}
\omega_\textrm{T}=\dfrac{\omega_1\omega_2(x_1+2x_2)}{2x_2\omega_1+x_1\omega_2},
\label{eq:125}  
\end{equation}
where $\omega_i$ and $x_i$ ($i=1,2$) are the resonant frequencies and reactance slope parameters of the individual resonators, respectively. If the linewidths of the resonances are equal (i.e., $x_2=x_1$), equation (\ref{eq:125}) is reduced to
\begin{equation}
\omega_\textrm{T}=\dfrac{3\omega_1\omega_2}{2\omega_1+\omega_2}.
\label{eq:126}
\end{equation}
From this equation, one can conclude that the frequency $\omega_\textrm{T}$ always lies between $\omega_1$ and $\omega_2$ since ${\cal X}_1$ and ${\cal X}_2$ have different signs only on this interval and, therefore, equation (\ref{eq:124}) can be satisfied. For the case $\omega_2=\omega_1$, the condition $\omega_\textrm{T}=\omega_2$ holds as expected. The normalized spectral spacing  between $\omega_\textrm{T}$ and $\omega_2$ is
\begin{equation}
\dfrac{\omega_\textrm{T}-\omega_2}{\omega_2}=\dfrac{\omega_1-\omega_2}{2\omega_1+\omega_2}\approx \dfrac{\omega_1-\omega_2}{3\omega_2}.
\label{eq:127}
\end{equation}
One can see from equation (\ref{eq:127}) that the higher perturbation (i.e., the higher difference between $\omega_1$ and $\omega_2$), the more spacing between frequencies $\omega_\textrm{T}$ and $\omega_2$. 

\section*{References}

\bibliographystyle{vancouver}
\bibliography{trimer}

\end{document}